\documentclass[pre,twocolumn,groupaddress,showpacs,showkeys,preprintnumbers,floatfix]{revtex4-1}

\usepackage{etex}
\usepackage[mathscr]{eucal}
\usepackage{ifpdf}
\usepackage[pagebackref]{hyperref}
\usepackage{dcolumn}
\usepackage{url}
\usepackage{amsmath}
\usepackage{amscd}
\usepackage{amsfonts}
\usepackage{amssymb}
\usepackage{amsthm}
\usepackage{bigints}
\usepackage{xfrac}
\usepackage{braket}
\usepackage{bm}   
\usepackage{bbm}
\usepackage[abs]{overpic}
\usepackage{verbatim}
\usepackage{stmaryrd}
\usepackage{xcolor}
\usepackage{setspace}
\usepackage{multirow}
\usepackage{tikz}
\usetikzlibrary{arrows}
\usetikzlibrary{automata}
\usetikzlibrary{positioning}
\usetikzlibrary{plotmarks}
\usepackage{pgfplots}
\pgfplotsset{compat=newest}

\usepackage{graphicx}
\usepackage[caption=false,justification=raggedright]{subfig}

\usepackage{hyphenat}
\tikzstyle{vaucanson}=[
  node distance=2.5cm,
  bend angle=15,
  auto,
  every loop/.style={},
  every edge/.style={->,draw=black,line width=1.2,>=latex,shorten <=1pt,shorten >=1pt},
  every state/.style={draw=black,line width=2,font=\large},
  loop right/.style={right,out=22,in=-22,loop},
  loop above/.style={above,out=112,in=68,loop},
  loop left/.style={left,out=202,in=158,loop},
  loop below/.style={below,out=292,in=248,loop},
  loop above right/.style={right,out=67,in=23,loop},
  loop above left/.style={left,out=157,in=113,loop},
  loop below left/.style={left,out=247,in=203,loop},
  loop below right/.style={right,out=337,in=293,loop},
]

\theoremstyle{plain}    
\theoremstyle{plain}    
\theoremstyle{plain}    
\theoremstyle{plain}    
\theoremstyle{plain}    
\theoremstyle{plain}    
\theoremstyle{plain}    
\theoremstyle{plain}    
\theoremstyle{plain}    
\theoremstyle{plain}    
\theoremstyle{plain}    \newtheorem{Def}{Definition}
\theoremstyle{plain}    
\theoremstyle{plain}

\DeclareMathOperator*{\argmax}{argmax}

\newcommand{\eM}     {\mbox{$\epsilon$-machine}}
\newcommand{\eMs}    {\mbox{$\epsilon$-machines}}
\newcommand{\EM}     {\mbox{$\epsilon$-Machine}}

\newcommand{\MeasAlphabet}  {\mathcal{A}}
\newcommand{\MeasSymbol}   { {X} }
\newcommand{\meassymbol}   { {x} }

\newcommand{\Past} { \smash{\overleftarrow {\MeasSymbol}} }

\newcommand{\Future}   { \smash{\overrightarrow{\MeasSymbol}} }

\newcommand{\CausalState}   { \mathcal{S} }

\newcommand{\causalstate}   { \sigma }
\newcommand{\CausalStateSet}    { \boldsymbol{\CausalState} }
\newcommand{\AlternateState}    { \mathcal{R} }

\newcommand{\AlternateStateSet} { \boldsymbol{\AlternateState} }

\newcommand{\Cmu}       {C_\mu}
\newcommand{\hmu}       {h_\mu}
\newcommand{\EE}        {{\bf E}}
\newcommand{\TI}        {{\bf T}}
\newcommand{\SI}        {{\bf S}}

\newcommand{\PC}        {\chi}

\newcommand{\ProcessAlphabet}   {\MeasAlphabet}

\newcommand{\forward}{+}
\newcommand{\reverse}{-}
\newcommand{\forwardreverse}{\pm} 

\newcommand{\FutureCausalState} { {\CausalState}^{\forward} }

\newcommand{\PastCausalState}   { {\CausalState}^{\reverse} }

\newcommand{\one}{\mathbf{1}}

\newcommand{\lastindex}[2]{
  \edef\tempa{0}
  \edef\tempb{#2}
  \ifx\tempa\tempb
    \edef\tempc{#1}
  \else
    \edef\tempa{0}
    \edef\tempb{#1}
    \ifx\tempa\tempb
      \edef\tempc{#2}
    \else
      \edef\tempc{#1+#2}
    \fi
  \fi
  \tempc
}

\newcommand{\bmu}{b_\mu}

\newcommand{\I}{\mathbf{I}}

\newcommand{\CSjoint}[1][,]{
   \edef\tempa{:}
   \edef\tempb{#1}
   \ifx\tempa\tempb
      \ensuremath{\FutureCausalState\!#1\PastCausalState}
   \else
      \ensuremath{\FutureCausalState#1\PastCausalState}
   \fi
}

\newif\ifpm
\edef\tempa{\forwardreverse}
\edef\tempb{\pm}
\ifx\tempa\tempb
   \pmfalse
\else
   \pmtrue
\fi

\renewcommand{\H}{\operatorname{H}}
\renewcommand{\I}{\operatorname{I}}
\newcommand{\EEsp}{\mathcal{E} (\omega)}

\newcommand{\Psp}{P(\omega)}

\colorlet {R_color}    {blue}
\colorlet {k_color}    {black!30!green}

\def\clap#1{\hbox to 0pt{\hss#1\hss}}

\begin{document}

\title{Spectral Simplicity of Apparent Complexity, Part II:\\
Exact Complexities and Complexity Spectra}

\author{Paul M. Riechers}
\email{pmriechers@ucdavis.edu}

\author{James P. Crutchfield}
\email{chaos@ucdavis.edu}

\affiliation{Complexity Sciences Center\\
Department of Physics\\
University of California at Davis\\
One Shields Avenue, Davis, CA 95616}

\date{\today}
\bibliographystyle{unsrt}

\begin{abstract}
The meromorphic functional calculus developed in Part I overcomes the
nondiagonalizability of linear operators that arises often in the
temporal evolution of complex systems and is generic to the metadynamics of
predicting their behavior. Using the resulting spectral decomposition, we derive closed-form
expressions for correlation functions, finite-length Shannon entropy-rate
approximates, asymptotic entropy rate, excess entropy, transient information,
transient and asymptotic state uncertainty, and synchronization information of
stochastic processes generated by finite-state hidden Markov models. This
introduces analytical tractability to investigating information processing in
discrete-event stochastic processes, symbolic dynamics, and chaotic dynamical
systems. Comparisons reveal mathematical similarities between complexity
measures originally thought to capture distinct informational and computational
properties. We also introduce a new kind of spectral analysis via coronal
spectrograms and the frequency-dependent spectra of past-future mutual
information. We analyze
a number of examples to illustrate the methods, emphasizing processes with
multivariate dependencies beyond pairwise correlation. An appendix presents
spectral decomposition calculations for one example in full detail.
\end{abstract}

\keywords{hidden Markov model, entropy rate, excess entropy, predictable
information, statistical complexity, projection operator, complex analysis,
resolvent, Drazin inverse}

\pacs{
02.50.-r  
89.70.+c  
05.45.Tp  
02.50.Ey  
02.50.Ga  
}
\preprint{Santa Fe Institute Working Paper 2017-06-XXX}
\preprint{arxiv.org:1706.XXXXX [nlin.cd]}

\maketitle


\setstretch{1.0}
\tableofcontents  

\setstretch{1.1}

\newcommand{\Abet}{\ProcessAlphabet}
\newcommand{\MS}{\MeasSymbol}
\newcommand{\ms}{\meassymbol}
\newcommand{\SSet}{\CausalStateSet}
\newcommand{\St}{\CausalState}
\newcommand{\st}{s}
\newcommand{\cs}{\causalstate}
\newcommand{\syncMSP}{{$\mathscr{S}$-MSP}}
\newcommand{\crypticMSP}{{$\PC$-MSP}}
\newcommand{\MxSt}{\AlternateState}
\newcommand{\MxSSet}{\AlternateStateSet_\pi}
\newcommand{\MxFSet}{\AlternateStateSet_\one}
\newcommand{\mxst}{\eta}
\newcommand{\mxstw}[1]{\mxst_{#1}} 		
\newcommand{\StartMS}{\bra{\delta_\pi}}
\newcommand{\opGen}{A}         
\newcommand{\matHMM}{T}     
\newcommand{\matMSP}{W}     
\newcommand{\TentT}{\varTheta}

\newcommand{\HWA}{\ket{\H(W^\Abet)}}
\newcommand{\Hmxst}{\ket{\H[\mxst]}}
\newcommand{\hsym}{\reflectbox{h}\text{h}}
\newcommand{\Redundancy}{\boldsymbol{R}}
\newcommand{\ACEphemeral}{\gamma_{\multimap}}
\newcommand{\hEphemeral}{h_{\multimap}}
\newcommand{\ACPersistent}{\gamma_{\rightsquigarrow}}
\newcommand{\hPersistent}{h_{\rightsquigarrow}}
\newcommand{\EEEphemeral}{\EE_{\multimap}}
\newcommand{\EEPersistent}{\EE_{\rightsquigarrow}}
\newcommand{\PU}{\reflectbox{$\mathcal{H}$}}

\newcommand{\kB}{k_\text{B}}
\newcommand{\corrbra}{\bra{\pi \overline{\Abet}}}
\newcommand{\corrket}{\ket{\Abet \one}}

\newcommand{\Cmatrix}{\mathcal{C}}   
\newcommand{\LWwoutZero}{\Lambda_W^{\setminus 0}}

\vspace{.3in}
{\bf
The prequel laid out a new toolset that allows one to analyze in detail how
complex systems store and process information. Here, we use the tools to
calculate in closed form almost all complexity measures for processes generated
by finite-state hidden Markov models. Helpfully, the tools also give a detailed
view of how subprocess components contribute to a process' informational
architecture. As an application, we show that the widely-used methods based on
Fourier analysis and power spectra fail to capture the structure of even very
simple structured processes. We introduce the spectrum of past-future mutual
information and show that it allows one to detect such structure.
}

\section{Introduction}

Tracking the evolution of a complex system, a time series of observations
often appears quite complicated in the sense of temporal patterns,
stochasticity, and behavior that require significant resources to predict.
Such complexity arises from many sources. Apparent complexity, even in simple
systems, can be induced by practical measurement and analysis issues, such as
small sample size, inadequate collection of probes, noisy or systematically
distorted measurements, coarse-graining, out-of-class modeling, nonconvergent
inference algorithms, and so on. The effects can either increase or decrease
apparent complexity, as they add or discard information, hiding the system of
interest from an observer to one degree or another. Assuming perfect observation, complexity can also be
inherent in nonlinear stochastic dynamical processes---deterministic chaos,
superexponential transients, high state-space dimension, nonergodicity,
nonstationarity, and the like. Even in ideal settings, the smallest sufficient
set of a system's maximally predictive features is generically uncountable,
making approximations unavoidable, in principle \cite{Marz17a}. With nothing
else said, these facts obviate physical science's most basic
goal---prediction---and, without that, they preclude understanding how nature
works. How can we make progress?

The prequel, Part I, argued that this is too pessimistic a view. It
introduced constructive results that address hidden structure and the
challenges associated with predicting complex systems. Part I showed that
questions regarding correlation, predictability, and prediction each require
their own analytical structures, as long as one can identify a system's hidden
linear dynamic. It distinguished two genres of quantitative question: (i)
\emph{cascading}, in which the influence of an initial preparation cascades
through state-space as time evolves, affecting the final measurement, and (ii)
\emph{accumulating}, in which statistics are gathered during such cascades. Part
I identified the linear algebraic structure underlying each kind.

Part I explained that the hidden linear dynamic in systems \emph{induces} a
nondiagonalizable metadynamics, even if the dynamics are diagonalizable in
their underlying state-space. Assuming normal and diagonalizable dynamics, so
familiar in mathematical physics, simply fails in this setting. Thus,
nondiagonalizable dynamics present an analytical roadblock. Part I reviewed a
calculus for functions of nondiagonalizable operators---the recently developed
meromorphic functional calculus of Ref.~\cite{Riec16a}---that directly addresses
nondiagonalizability, giving constructive calculational methods and algorithms.

Along the way, Part I reviewed relevant background in stochastic processes and
their complexities and the hidden Markov models (HMMs) that generate them. It
delineated several classes of HMMs---Markov chains, unifilar HMMs, and
nonunifilar HMMs. It also reviewed their mixed-state presentations (MSPs)---HMM
generators of a process that track distributions induced by observation.  Related constructions
included the HMM and \eM\ synchronizing MSPs, generator mixed-functional
presentations, and cryptic-operator presentations. MSPs are key to calculating
complexity measures within an information-theoretic framing. Part I then showed
how each complexity measure reduces to a linear algebra of an appropriate HMM
adapted to the cascading- or accumulating-question genre. It summarized the
meromorphic functional calculus and several of its mathematical implications in
relation to projection operators. Part I also highlighted a spectral weighted
directed-graph theory that can give useful shortcuts for determining a process'
spectral decomposition. Part II here uses Part I's notation and assumes
familiarity with its results.

With Part I's toolset laid out, Part II now derives the promised closed-form
complexities of a process. Section \S \ref{sec:TransientBehavior} investigates
the range of possible behaviors for correlation and myopic uncertainty via
convergence to asymptotic correlation and asymptotic entropy rates. Section \S
\ref{sec:DiagAccumulation} then considers measures related to accumulating
quantities during the transient relaxation to synchronization. Section \S
\ref{sec:ExactComplexity} introduces closed-form expressions for a wide range
of complexity measures in terms of the spectral decomposition of a process'
dynamic. It also introduces complexity spectra and highlights common
simplifications for special cases, such as almost diagonalizable dynamics.
Section \S \ref{sec:CoronalSpec} gives a new kind of signal analysis in terms
of coronal spectrograms. A suite of examples in \S \ref{sec:Examples} and \S
\ref{sec:RRX} ground the theoretical developments and are complemented with an
in-depth pedagogical example worked out in App. \S \ref{sec:ExactCalculation}.
Finally, we conclude with a brief retrospective of Parts I and II and give an
eye towards future applications.

\section{Correlation and Myopic Uncertainty}
\label{sec:TransientBehavior}

Using Part I's methods, our first step is to solve for the
\emph{correlation function}:
\begin{align}
\gamma(L) & = \left\langle \overline{\MS}_t \MS_{t+L} \right\rangle_t
\label{eq:CorrFunc}
\end{align}
and the \emph{myopic uncertainty} or finite-history Shannon entropy rate:
\begin{align}
\hmu(L) &= 
\H \left[ X_{L} \middle| X_{1 : L} \right]
  ~.
\label{eq:MyopicUncertainty}
\end{align}
A comparison is informative. We then determine the asymptotic correlation and
myopic uncertainty from the resulting finite-$L$ expressions.

\subsection{Nonasymptotics}
\label{sec:Nonasymptotics}

A central result in Part I was the spectral decomposition of powers of a linear
operator $A$, even if that operator is nondiagonalizable. Recall that for any
$L \in \mathbb{C}$:
\begin{align}
\opGen^L
    & = \Biggl[ \sum_{\lambda \in \Lambda_\opGen \atop \lambda \neq 0}
\sum_{m  = 0}^{\nu_\lambda - 1}
         \binom{L}{m} \lambda^{L-m}
         \opGen_{\lambda, m} \Biggr] \nonumber \\
         & \qquad + \left[ 0 \in \Lambda_\opGen \right]
          \sum_{m = 0}^{\nu_0 - 1}
          \delta_{L, m}  \opGen_0  \opGen^m
          ~,
\label{eq: T^n generally}
\end{align}
where $\binom{L}{m}$ is the generalized binomial coefficient:
\begin{align}
\binom{L}{m} & = \frac{1}{m!} \prod_{n=1}^m (L-n+1)
  ~,
\end{align}
$\binom{L}{0} = 1$, and $[ 0 \in \Lambda_\opGen ]$ is the \emph{Iverson
bracket}. The latter takes on value $1$ if zero is an eigenvalue of $\opGen$
and $0$ if not.

In light of this, the autocorrelation function $\gamma(L)$ is simply a superposition of weighted eigen-contributions. Part I showed that Eq. (\ref{eq:CorrFunc}) has the operator expression:
\begin{align*}
\gamma(L) = \corrbra T^{|L|-1} \corrket
  ~,
\end{align*}
where $T$ is the transition dynamic, $\Abet$ is the output symbol alphabet, and we defined the row vector:
\begin{align*}
\corrbra & = \bra{\pi} \Big( \sum_{\ms \in \Abet} \overline{\ms} T^{(\ms)} \Big)
\end{align*}
and the column vector:
\begin{align*}
 \corrket & = \Big( \sum_{\ms \in \Abet}  \ms T^{(\ms)} \Big) \ket{\one}
  ~.
\end{align*}
Substituting Part I's spectral decomposition of matrix powers, Eq. (\ref{eq:
T^n generally}) above, directly leads to the spectral decomposition of
$\gamma(L)$ for nonzero integer $L$:
\begin{align} 
\gamma(L) & = \sum_{\lambda \in \Lambda_T \atop \lambda \neq 0} 
  \sum_{m = 0}^{\nu_\lambda -1} \corrbra  T_{\lambda, m} \corrket
   \binom{| L |-1}{m}  \lambda^{| L |-1-m} \nonumber \\
  & \qquad + [0 \in \Lambda_{T}] 
   \sum_{m = 0}^{\nu_0 -1} \corrbra  T_{0} T^{m}  \corrket \delta_{| L |-1, m}
\label{eq:AutocorrelationDecomposed} \\
  & = \ACPersistent(L) + \ACEphemeral(L)
  ~.
\end{align} 
We denote the persistent first term of Eq.~\eqref{eq:AutocorrelationDecomposed}
as $\ACPersistent$, and note that it can be expressed:
\begin{align*}
\ACPersistent (L) & = \corrbra T^\mathcal{D} T T^{|L|-1} \corrket \\
  & = \corrbra T^\mathcal{D} T^{|L|} \corrket
  ~,
\end{align*}
where $T^\mathcal{D}$ is $T$'s Drazin inverse. We denote the ephemeral second
term as $\ACEphemeral$, which can be written as:
\begin{align*}
\ACEphemeral (L) = \corrbra T_0 T^{|L| - 1} \corrket
  ~,
\end{align*}
where $T_0$ is the eigenprojector associated with the eigenvalue of zero;
$T_0 = \boldsymbol{0}$ if $0 \notin \Lambda_T$.

From Eq.~\eqref{eq:AutocorrelationDecomposed}, it is now apparent that the
index of $T$'s zero eigenvalue gives a finite-horizon contribution
($\ACEphemeral$) to the autocorrelation function. Beyond index $\nu_0$ of $T$,
the only $L$-dependence comes via a weighted sum of terms of the form
$\binom{|L|-1}{m} \lambda^{|L|-1-m}$---polynomials in $L$ times decaying
exponentials. The set $\bigl\{ \corrbra  T_{\lambda, m} \corrket \bigr\}$
simply weights the amplitudes of these contributions. In the familiar
diagonalizable case, the behavior of autocorrelation is simply a sum of
decaying exponentials $\lambda^{|L|}$.

Similarly, in light of Part I's expression for the myopic entropy rate in terms
of the MSP---starting in the initial unsynchronized mixed-state $\pi$ and
evolving the state of uncertainty via the observation-induced MSP transition
dynamic $W$:
\begin{align}
\hmu(L) &= 
\StartMS W^{L-1} \HWA
\label{eq:MyopticUncertainty}
\end{align}
---and its spectral
decomposition of $A^L$, we find the most general spectral decomposition of the
myopic entropy rates $\hmu(L)$ to be:
\begin{align} 
\hmu(L) & = \sum_{\lambda \in \Lambda_W \atop \lambda \neq 0}
  \sum_{m = 0}^{\nu_\lambda -1} \StartMS W_{\lambda, m} \HWA
  \binom{L-1}{m} \lambda^{L-1-m} \nonumber \\ 
  &  +  \left[ 0 \in \Lambda_W \right] \sum_{m=0}^{\nu_0 - 1} \delta_{L-1, m} 
	\StartMS W_0 W^m \HWA 
	\label{eq:hmuLdecomposed} \\
  & = 	\hPersistent(L) + \hEphemeral(L)
  ~.
\label{eq:2kindsOfhmuL}
\end{align} 
We denote the persistent first term of Eq. (\ref{eq:hmuLdecomposed}) as
$\hPersistent$, and note that it can be expressed directly as:
\begin{align*}
\hPersistent (L) & = \StartMS W^\mathcal{D} W W^{L-1} \HWA \\
                 & = \StartMS W^\mathcal{D} W^{L} \HWA
  ~,
\end{align*}
where $W^\mathcal{D}$ is the Drazin inverse of the mixed-state-to-state net transition dynamic $W$. We denote the ephemeral second term as $\hEphemeral$, 
which can be written as:
\begin{align*}
\hEphemeral (L) = \StartMS W_0 W^{L - 1} \HWA
  ~.
\end{align*}

From Eq.~\eqref{eq:hmuLdecomposed}, we see that the index of $W$'s zero
eigenvalue gives a finite horizon contribution ($\hEphemeral$) to the myopic
entropy rate. Beyond index $\nu_0$ of $W$, the only $L$-dependence comes via a
weighted sum of terms of the form $\binom{L-1}{m}
\lambda^{L-1-m}$---polynomials in $L$ times decaying exponentials. The set
$\bigl\{ \StartMS  W_{\lambda, m} \HWA \bigr\}$ weights the amplitudes of
these contributions.

For stationary processes we anticipate that, for all $\zeta \in \{ \lambda \in
\Lambda_W : |\lambda| = 1, \lambda \neq 1 \}$, $\StartMS W_{\zeta} =
\mathbf{0}$ and thus $\StartMS W_{\zeta} \HWA = 0$. Hence, we can save
ourselves from superfluous calculation by excluding the nonunity eigenvalues on
the unit circle, when calculating the myopic entropy rate for stationary
processes.  In the diagonalizable case, again, its behavior is simply a sum of
decaying exponentials $\lambda^{L}$.

In practice, $\ACEphemeral$ often vanishes, whereas $\hEphemeral$ is often
nonzero. This practical difference between $\ACEphemeral$ and $\hEphemeral$
stems from the difference between typical graph structures of the respective
dynamics. For a stationary process' generic transition dynamic, zero
eigenvalues (and so $\nu_0(T)$ of $T$) typically arise from hidden symmetries
in the dynamic. In contrast, the MSP of a generic transition dynamic often has
tree-like ephemeral structures that are primarily responsible for the zero
eigenvalues (and $\nu_0(W)$). Nevertheless, despite their practical typical
differences, the same mathematical structures appear and contribute to the most
general behavior of each of these cascading quantities.

The breadth of qualitative behaviors shared by autocorrelation and myopic
entropy rate is common to the solution of all questions that can be
reformulated as a cascading hidden linear dynamic; the myopic state uncertainty
$\mathcal{H}^{\forward}(L)$ is just one of many other examples. As we have
already seen, however, different measures of a process reflect signatures of
different linear operators.

Next, we explore similarities in the qualitative behavior of asymptotics
and discuss the implications for correlation and entropy rate.

\subsection{Asymptotic correlation}

The spectral decomposition reveals that the autocorrelation converges to a
constant value as $L \to \infty$, unless $T$ has eigenvalues on the unit circle
besides unity itself. This holds if index $\nu_0$ is finite, which it is for
all processes generated by finite-state HMMs and also many infinite-state HMMs.
If unity is the sole eigenvalue with magnitude one, then all other
eigenvalues have magnitude less than unity and their contributions vanish for
large enough $L$. Explicitly, if $\argmax_{\lambda \in \Lambda_{T}} |\lambda| =
\{ 1 \}$, then:
\begin{align*}
\lim_{L \to \infty} & \gamma(L)  \\
& = \lim_{L \to \infty} 
    \sum_{\lambda \in \Lambda_T \atop \lambda \neq 0} 
    \sum_{m = 0}^{\nu_\lambda -1} 
       \corrbra  T_{\lambda, m} \corrket
       \binom{L -1}{m}  \lambda^{L-1-m} \\
& = \corrbra  T_{1} \corrket \\ 
& = \corrbra  \one \rangle \langle \pi \corrket \\ 
&= \Bigl| \sum_{\ms \in \Abet} \ms \Pr(\ms) \Bigr|^2 \\
& = \bigl| \braket{x} \bigr|^2
  ~.
\end{align*}
This used the fact that $\nu_1 = 1$ and that $T_1 = \ket{\one} \bra{\pi}$ for
an ergodic process.

If other eigenvalues in $\Lambda_{T}$ besides unity lie on the unit circle, 
then the autocorrelation approaches a periodic sequence as $L$ gets large. 

\subsection{Asymptotic entropy rate}

By the Perron--Frobenius theorem, $\nu_\lambda = 1$ for all eigenvalues of $W$
on the unit circle. Hence, in the limit of $L \to \infty$, we obtain the asymptotic entropy rate for any stationary process: 
\begin{align}  
\hmu & \equiv \lim_{L \to \infty} h_\mu(L) \\ 
  & = \lim_{L \to \infty} 
\label{eq: sum for hu with only unity mag eigs}
  \sum_{\lambda \in \Lambda_W \atop  |\lambda| = 1 } \lambda^{L-1}
  \StartMS W_{\lambda} \HWA \\  
  & = 
\label{eq: hu with only W1 operator}
  \StartMS W_{1} \HWA 
  ~,
\end{align} 
since, for stationary processes, $\StartMS W_{\zeta}  = \mathbf{0}$ for all
$\zeta \in \{ \lambda \in \Lambda_W : |\lambda| = 1, \lambda \neq 1 \}$. For
nonstationary processes, the limit may not exist, but $\hmu$ may still be found
in a suitable sense as a function of time. If the process has only one
stationary distribution over mixed states, then $W_1 = \ket{\one}\bra{\pi_W}$
and we have:
\begin{align}  
\hmu & = 
\label{eq: hu as braket involving msp}
  \langle \pi_W \HWA
  ~,  
\end{align} 
where $\pi_W$ is the stationary distribution over $W$'s states,
found either from $\bra{\pi_W} = \StartMS W_{1}$ or from
solving $\bra{\pi_W} W = \bra{\pi_W}$.

A simple but interesting example of when ergodicity does \emph{not} hold is the
multi-armed bandit problem~\cite{Cover70,Crut15a}. In this, a realization is
drawn from an ensemble of differently biased coins or, for that matter, over
any other collection of IID processes. More generally, there can be many
distinct memoryful stationary components from which a given realization is
sampled, according to some probability distribution. With many attracting components we
have the stationary mixed-state eigenprojector $W_1 = \sum_{k=1}^{a_1}
\ket{1_k} \bra{1_k} $, with $\braket{1_j | 1_k} = \delta_{j,k}$, where the algebraic multiplicity $a_1(T) = a_1(W)$ of the `1' eigenvalue is the number of attracting components. The entropy rate becomes:
\begin{align}
\hmu &= \sum_{k=1}^{a_1}  \StartMS 1_k \rangle \langle 1_k \HWA \\
  & = \bigl\langle \hmu^{(\text{component } k)} \bigr\rangle_k
  ~.
\end{align}
Above, $\StartMS 1_k \rangle$ is the probability of ending up in component $k$,
while $\langle 1_k \HWA$ is component $k$'s entropy rate. Thus, if nonergodic,
the \emph{process'} entropy rate may not be the same as the entropy of any
particular realization. Rather, the process' entropy rate is a weighted average
of those for the ensemble of sequences constituting the process.

For unifilar $\mathcal{M}$, the topology, transition probabilities, and stationary distribution over the recurrent states are the same for both $\mathcal{M}$ and its \syncMSP. Hence, for unifilar $\mathcal{M}$ we have:
\begin{align}  
h_\mu & = \langle \pi_W \HWA \nonumber \\ 
  & = \braket{ \pi | H(T^\Abet)}
  ~. 
\label{eq: hu for unifilar M in braket form}
\end{align} 
One can easily show that Eq.\ \eqref{eq: hu for unifilar M in braket form} is equivalent to the well-known closed-form expression for $h_\mu$ for unifilar presentations:
\begin{align}  
\braket{ \pi | H(T^\Abet)}
  = - \sum_{\cs \in \SSet} \Pr(\cs)
  \sum_{\substack{\ms \in \Abet \\ \cs' \in \mathcal{S} }}
  T^{(\ms)}_{\cs, \cs'}
  \log_2  (T^{(\ms)}_{\cs, \cs'})
  ~. 
\label{eq: equivalence of hu expression to well known form}
\end{align}
For nonunifilar presentations, however, we must use the more general result of
Eq. \eqref{eq: hu as braket involving msp}. This is similar to the calculation
in Eq. \eqref{eq: equivalence of hu expression to well known form}, but must be
performed over the recurrent states of a mixed-state presentation, which may be
countable or uncountable.

\section{Accumulated Transients for Diagonalizable Dynamics}
\label{sec:DiagAccumulation}

In the diagonalizable case, autocorrelation, myopic entropy rate, and myopic
state uncertainty reduce to a sum of decaying exponentials. Correspondingly, we
can find the power spectrum, excess entropy, and synchronization information
respectively via geometric progressions.

For example, if $W$ is diagonalizable and has no zero eigenvalue, then the
myopic entropy rate reduces to:
\begin{align*}
\hmu(L) & = \sum_{\lambda \in \Lambda_W} 
  \StartMS W_{\lambda} \HWA  \lambda^{L-1}  \\
  & = \StartMS W_{1} \HWA
  + \!\! \sum_{\lambda \in \Lambda_W \atop  |\lambda| < 1 }
  \lambda^{L-1} \StartMS W_{\lambda} \HWA
  ,
\end{align*}
where $\StartMS W_{1} \HWA $ is identifiable as the entropy rate $\hmu$.

It then follows that the \emph{excess entropy}, which is the mutual information $\EE = \I[\Past; \Future]$ between the past and the future, is:
\begin{align} 
\EE & \equiv 
  \sum_{L=1}^\infty \left[ h_\mu(L) - h_\mu \right] \nonumber \\ 
  & = \sum_{L=1}^\infty  
  \sum_{\lambda \in \Lambda_W \atop  |\lambda| < 1 } \lambda^{L-1} 
  \StartMS W_{\lambda}  \HWA  \nonumber \\ 
  & = \sum_{\lambda \in \Lambda_W \atop  |\lambda| < 1 } 
  \StartMS W_{\lambda}  \HWA
  \!\!\!\!
  \underbrace{ \sum_{L=1}^\infty \lambda^{L-1} }_{= \sum_{L=0}^\infty
  \lambda^{L} = \frac{1}{1 - \lambda} }  \nonumber \\ 
  & = \sum_{\lambda \in \Lambda_W \atop  |\lambda| < 1 }
  \frac{1}{1 - \lambda} 
  \StartMS  W_{\lambda}  \HWA   
  ~.
\label{eq:EEFormulaDiag}
\end{align}
Note that larger eigenvalues (closer to unity magnitude) drive the denominator
$1 - \lambda$ closer to zero and, thus, increase $\frac{1}{1 - \lambda}$.
Hence, larger eigenvalues---controlling modes of the mixed-state transition
matrix that decay slowly---have the potential to contribute most to excess
entropy. Small eigenvalues---quickly decaying modes---do not contribute.
Putting aside the language of eigenvalues, one can paraphrase:
slowly decaying transient behavior (of the distribution of distributions over
process states) has the most potential to make a process appear complex.

Continuing, the \emph{transient information} is:
\begin{align*} 
\TI & \equiv 
  \sum_{L=1}^\infty L \left[ h_\mu(L) - h_\mu \right] \\ 
  & = \sum_{L=1}^\infty  
  \sum_{\lambda \in \Lambda_W \atop  |\lambda| < 1 } L \lambda^{L-1} 
  \StartMS W_{\lambda}  \HWA  \\ 
  & = \sum_{\lambda \in \Lambda_W \atop  |\lambda| < 1 } 
  \StartMS  W_{\lambda}  \HWA  
\!\!\!\!\!\!\!\!\!\!\!\!\!\!\!\!\!\!\!\!\!\!\!\!\!\!\!\!\!\!\!\!\!\!\!\!\!\!\!  \underbrace{ \sum_{L=1}^\infty L \lambda^{L-1} }_{ 
= \sum_{L=0}^\infty \frac{d}{d\lambda} \lambda^{L} 
= \frac{d}{d\lambda} \left(  \sum_{L=0}^\infty  \lambda^{L}   \right) 
= \frac{d}{d\lambda} \left(   \frac{1}{1 - \lambda}    \right) 
= \frac{1}{(1 - \lambda)^2} }  \\ 
  & = \sum_{\lambda \in \Lambda_W \atop  |\lambda| < 1 } 
  \frac{1}{(1 - \lambda)^2} \StartMS  W_{\lambda}  \HWA   
  ~.
\end{align*}
We now see that the transient information is very closely related to the excess
entropy, differing only via the square in the denominators. This comparison
between $\EE$ and $\TI$ closed-form expressions suggests an entire hierarchy of
informational quantities based on eigenvalue weighting.

Performing a similar procedure for the \emph{synchronization information}
$\SI'$ shows that:
\begin{align} 
\SI' & \equiv \sum_{L=0}^\infty  \bigl[ \mathcal{H}(L) - \mathcal{H} \bigr]  \nonumber \\
  & = \sum_{L=0}^\infty \sum_{\lambda \in \Lambda_W \atop  |\lambda| < 1 } \!
  \StartMS  W_{\lambda} \Hmxst  \, \lambda^{L}
  \nonumber \\  
  & = \sum_{\lambda \in \Lambda_W \atop  |\lambda| < 1 }
  \StartMS  W_{\lambda} \Hmxst  \sum_{L=0}^\infty  \lambda^{L} \nonumber \\ 
  & = 
  \sum_{\lambda \in \Lambda_W \atop  |\lambda| < 1 } 
  \frac{1}{1 - \lambda} 
  \StartMS  W_{\lambda} \Hmxst 
  ~.\nonumber  
\end{align} 

The expressions reveal a remarkably close relationship between $\SI'$ and $\EE$. 
Define $\bra{\cdot} \equiv \sum_{L=0}^\infty \StartMS W^{L}$. Then:
\begin{align*}
\bra{\cdot} = \sum_{\lambda \in \Lambda_W \atop  |\lambda| < 1 } 
  \frac{1}{1 - \lambda} \StartMS  W_{\lambda}
  ~.
\end{align*}
The relationship is now made plain:
\begin{align*}
\EE &= \langle \cdot \HWA \text{~~and} \\  
\SI' &= \langle \cdot \Hmxst
  ~.  
\end{align*}
Although a bit more cumbersome, perhaps better intuition emerges if we rewrite
$\bra{\cdot}$ as $\bra{\int \Pr(\mxst, L) dL}$.

Again, large eigenvalues---slowly decaying modes of the mixed-state transition
matrix---can make the largest contribution to synchronization information;
small eigenvalues correspond to quickly decaying modes that do not have the
opportunity to contribute. In fact, the potential of large eigenvalues to make
large contributions is a recurring theme for many questions one has about a
process. Simply stated, long-term behavior---what we often interpret as
``complex'' behavior---is dominated by a process's largest-eigenvalue modes.

That said, a word of warning is in order. Although large-eigenvalue modes have
the most \emph{potential} to make contributions to a process's complexity, the
actual set of largest contributors also depends strongly on the amplitudes $\{
\StartMS  W_{\lambda}  \ket{\dots} \}$, where $\ket{\dots}$ is some quantifier
vector of interest; e.g., $\ket{\dots} = \Hmxst$, $\ket{\dots} = \HWA$, or
$\ket{\dots} = \ket{\one}$.

Hence, there is as-yet unanticipated similarity between $\EE$ and $\TI$ and
another between $\EE$ and $\SI'$---at least assuming diagonalizability.  We
would like to know the relationships between these quantities more generally.
However, deriving the general closed-form expressions for accumulated
transients is not tractable via the current approach. Rather, to derive the
general results, we deploy the meromorphic functional calculus directly at an
elevated level, as we now demonstrate.

\section{Exact Complexities and Complexity Spectra}
\label{sec:ExactComplexity}

We now derive the most general closed-form solutions for several complexity
measures, from which expressions for related measures follow straightforwardly.
This includes an expression for the past--future mutual information or excess
entropy, identifying two distinct persistent and transient components, and a
novel extension of excess entropy to temporal frequency spectra components. We
also give expressions for the synchronization information and power spectra. We
explicitly address the class---a common one we argue---of almost diagonalizable
dynamics. The section finishes by highlighting finite-order Markov order
processes that, rather than being simpler than infinite Markov order processes,
introduce technical complications that must be addressed.

Before carrying this out, we define several useful objects. Let $\rho(A)$ be
the spectral radius of matrix $A$:
\begin{align*}
\rho(A) = \max_{\lambda \in \Lambda_A} | \lambda |
  ~.
\end{align*}
For stochastic $W$, since $\rho(W) = 1$, let $\Lambda_{\rho(W)}$ denote the set of eigenvalues with unity magnitude:
\begin{align*} 
\Lambda_{\rho(W)} = \{ \lambda \in \Lambda_W: | \lambda | = 1 \}
  ~.
\end{align*}

We also define:
\begin{align} 
Q \equiv W -  W_1 
\end{align} 
and 
\begin{align} 
\mathcal{Q} \equiv W - \sum_{\lambda \in \Lambda_{\rho(W)}} \lambda W_\lambda
  ~.
\end{align}

Eigenvalues with unity magnitude that are not themselves unity correspond to perfectly periodic cycles of the state-transition dynamic. By their very nature, such cycles are restricted to the recurrent states. Moreover, we expect the projection operators associated with these cycles to have no 
net overlap with the start-state of the MSP. So, we expect:
\begin{align}
\StartMS W_\lambda & = \bra{0}
  ~,
\label{eq:CyclesDontContribute}
\end{align} 
for all $\lambda \in \Lambda_{\rho(W)} \setminus \{ 1 \}$. Hence: 
\begin{align}
\StartMS Q^L & = \StartMS \mathcal{Q}^L
  ~. 
\end{align} 

We will also use the fact that, since $\rho(\mathcal{Q}) < 1$:
\begin{align*}
\sum_{L = 0}^\infty \mathcal{Q}^L &= (I - \mathcal{Q})^{-1} ~;
\end{align*} 
and furthermore:
\begin{align*}
\StartMS  (I - \mathcal{Q})^{-1} &= \StartMS  (I - Q)^{-1} 
  ~,
\end{align*} 
as a consequence of Eq. \eqref{eq:CyclesDontContribute} and our spectral
decomposition.

Having seen complexity measures associated with prediction all take on a
similar form in terms of the \syncMSP\ state-transition matrix, we expect to
encounter similar forms for generically nondiagonalizable state-transition
dynamics.

\subsection{Excess entropy}

We are now ready to develop the excess entropy in full generality.
Our tools turn this into a direct calculation. We find: 
\begin{align*}
\EE & \equiv \sum_{L=1}^\infty \left[ h_\mu(L) - h_\mu \right] \nonumber \\ 
  & = \sum_{L=1}^\infty
  \left[ \StartMS W^{L-1} \HWA - \StartMS W_1 \HWA \right] \\
  & = \sum_{L=0}^\infty
  \left[ \StartMS W^{L} \HWA - \StartMS W_1 \HWA \right] \\
  & = \sum_{L=0}^\infty  \StartMS
  \bigl[  (\underbrace{W - W_1}_{\equiv Q})^L - \delta_{L, 0} W_1 \bigr] \HWA \\
  & = - \underbrace{\StartMS W_1 \HWA}_{= \hmu  }
  + \sum_{L=0}^\infty
  \underbrace{ \StartMS  Q^L }_{= \StartMS \mathcal{Q}^L}  \HWA \\
  & =  \StartMS \Bigl( \sum_{L=0}^\infty \mathcal{Q}^L \Bigr) \HWA - \hmu \\ 
  & =  \StartMS ( I - \mathcal{Q})^{-1} \HWA - \hmu \\ 
  & =  \StartMS ( I - Q)^{-1} \HWA - \hmu
  ~. 
\end{align*}
Note that $(I-Q)^{-1} = \text{inv}(I-Q)$ here, since unity is not an eigenvalue
of $Q$. Indeed, the unity eigenvalue was explicitly extracted from the former
matrix to make an invertible expression.

For an ergodic process, where $W_1 = \ket{\one} \bra{\pi_W}$, this becomes:
\begin{align}
\EE & =  \StartMS
  \bigl( I - W + \ket{\one} \bra{\pi_W} \, \bigr)^{-1} \HWA - \hmu
  ~. 
\label{eq: useful EE eq}
\end{align}
Computationally, Eq.\ \eqref{eq: useful EE eq} is wonderfully useful. However,
the subtraction of $\hmu$ is at first mysterious. Especially so, when compared
to the compact result for the excess-entropy spectral decomposition in the
diagonalizable case given by Eq.~\eqref{eq:EEFormulaDiag}.

Let's explore this. Recall that Ref.~\cite{Riec16a} showed:
\begin{align}
(I - T)^{\mathcal{D}} & = \left[ I - (T - T_1) \right]^{-1} - T_1
\label{eq: Drazin vs I less Q}
  ~,
\end{align}
for any stochastic matrix $T$.
From this, we see that the general solution for $\EE$ takes on its most elegant form in terms of the Drazin inverse of $I-W$: 
\begin{align}
\EE & =  \StartMS ( I - Q)^{-1} \HWA - \hmu \nonumber \nonumber \\
  & =  \StartMS ( I - Q)^{-1} \HWA - \StartMS W_1 \HWA \nonumber \\ 
  & =  \StartMS \left[ ( I - Q)^{-1} - W_1 \right]  \HWA \nonumber \\ 
  & = \StartMS  ( I - W)^{\mathcal{D}}  \HWA
  ~.
\label{eq: EE in terms of Drazin}
\end{align}

Recall too Part I's explicit spectral decomposition:
\begin{align}
(I - T)^{\mathcal{D}} & =
 \!\!\! \sum_{\lambda \in \Lambda_T \setminus \{ 1 \} }
    \sum_{m=0}^{\nu_\lambda - 1}
    \frac{1}{(1-\lambda)^{m+1}}
    T_{\lambda, m} ~.
\label{eq:ResolventAt1}
\end{align}
From this and Eq.~\eqref{eq: EE in terms of Drazin}, we see that the
past--future mutual information---the amount of the future that is
\emph{predictable} from the past---has the general spectral decomposition:
\begin{align}
\EE = \!\!\! \sum_{\lambda \in \Lambda_W \setminus \{ 1 \} } 
    \! \sum_{m=0}^{\nu_\lambda - 1} 
    \frac{1}{(1-\lambda)^{m+1}} 
    \StartMS  W_{\lambda, m}  \HWA
  .
\end{align}

\subsection{Persistent excess}

In light of Eq.~\eqref{eq:2kindsOfhmuL}, we see that there are two qualitatively distinct contributions to the excess entropy $\EE = \EEPersistent + \EEEphemeral$.
One comprises the persistent leaky contributions from all $L$:
\begin{align*}
\EEPersistent
  & \equiv \sum_{L=1}^\infty \left[ \hPersistent(L) - h_\mu \right] \\
  & = \StartMS W^\mathcal{D} W (I - W)^\mathcal{D} \HWA 
\end{align*}
and the other is a completely ephemeral piece that contributes only up to $W$'s zero-eigenvalue index $\nu_0$:
\begin{align*}
\EEEphemeral & \equiv \sum_{L=1}^\infty \hEphemeral(L) \\
  & = \sum_{L=1}^{\nu_0} \hEphemeral(L) \\
  & = \StartMS W_0 (I - W)^\mathcal{D} \HWA
  ~.
\end{align*}

\subsection{Excess entropy spectrum}

Equation \eqref{eq: EE in terms of Drazin} immediately suggests that we
generalize the excess entropy, a scalar complexity measure, to a complexity
\emph{function} with continuous part defined in terms of the resolvent---say,
via introducing the complex variable $z$:
\begin{align*}
E(z) = \StartMS (zI-W)^{-1} \HWA
  ~.
\end{align*}
Such a function not only monitors \emph{how much} of the future is predictable,
but also reveals the \emph{time scales of interdependence} between the
predictable features within the observations. Directly taking the $z$-transform
of $\hmu(L)$ comes to mind, but this requires tracking both real and imaginary
parts or, alternatively, both magnitude and complex phase. To ameliorate this,
we employ a transform of a closely related function that contains the same
information.

Before doing so, we should briefly note that ambiguity surrounds the
appropriate excess-entropy generalization. There are many alternate measures
that approach the excess entropy as frequency goes to zero. For example,
directly calculating from the meromorphic functional calculus, letting $z =
e^{i \omega}$ we find:
\begin{align*}
\lim_{\omega \to 0} \text{Re} \StartMS (e^{i \omega} I - W)^{-1} \HWA
  = \EE - \frac{1}{2} \hmu
  ~.
\end{align*}
We are challenged, however, to interpret the fact that
$\text{Re} \StartMS (e^{i \omega} I - W)^{-1} \HWA + \frac{1}{2} \hmu$ is not
necessarily positive at all frequencies. Another direct calculation shows that:
\begin{align*}
\lim_{\omega \to 0} \text{Re} \StartMS e^{i \omega} (e^{i \omega} I - W)^{-1}
\HWA = \EE + \frac{1}{2} \hmu
  ~.
\end{align*}
Enticingly, $\text{Re} \StartMS e^{i \omega} (e^{i \omega} I - W)^{-1} \HWA -
\frac{1}{2} \hmu$ appears to be positive over all frequencies for all examples
checked. It is not immediately clear which, if either, is the appropriate
generalization, though. Fortunately, the Fourier transform of a two-sided
myopic-entropy convergence function makes our upcoming definition of $\EEsp$
interpretable and of interest in its own right.

Let $\hsym$ be the \emph{two-sided myopic entropy convergence function} defined
by:
\begin{align*}
\hsym (L) & = 
\begin{cases}
\H(\MS_0 | \MS_{-\left| L \right| +1:0}) & \text{for } L < 0 ~, \\
\log_2(\Abet) & \text{for } L = 0 ~, \text{and} \\
\H(\MS_0 | \MS_{1:L}) & \text{for } L > 0 ~.
\end{cases}
\end{align*}
For stationary processes, it is easy to show that $\H(\MS_0 | \MS_{-L+1:0}) =
\H(\MS_0 | \MS_{1:L})$, with the result that $\hsym$ is a symmetric function.
Moreover, $\hsym$ then simplifies to:
\begin{align*}
\hsym (L) & = \hmu(|L|) ~,
\end{align*}
where $\hmu(0) \equiv \log_2(\Abet) $ and, as before, $\hmu(L) = \H(\MS_L |
\MS_{1:L})$ for $L \geq 1$ with $\hmu(1) = \H(\MS_1)$.

The symmetry of the two-sided myopic entropy convergence function $\hsym$
guarantees that its Fourier transform is also real and symmetric. Explicitly,
the continuous part of the Fourier transform turns out to be:
\begin{align*}
\widetilde{\hsym}_\text{c}(\omega) = \Redundancy + 2 \text{Re} \StartMS (e^{i \omega} I - W)^{-1} \HWA ~,
\end{align*}
a strictly real and symmetric function of the angular frequency $\omega$.
Here, $\Redundancy$ is the redundancy of the alphabet $\Redundancy \equiv
\log_2 | \Abet | - \hmu$, as in Ref.~\cite{Crut01a}.

The transform $\widetilde{\hsym}$ also has a discrete impulsive component. For
stationary processes this consists solely of the Dirac delta function at zero
frequency:
\begin{align*}
\widetilde{\hsym}_\text{d}(\omega) = 2 \pi \hmu \sum_{k \in \mathbb{Z} } \delta(\omega + 2 \pi k)
  ~.
\end{align*}
Recall that the Fourier transform of a discrete-domain function is $2
\pi$-periodic in the angular frequency $\omega$. This delta function is
associated with the nonzero offset of the entropy convergence curve of
positive-entropy-rate processes. The full transform is:
\begin{align*}
\widetilde{\hsym}(\omega)
  = \widetilde{\hsym}_\text{c}(\omega) + \widetilde{\hsym}_\text{d}(\omega)
  ~.
\end{align*}

Direct calculation using the meromorphic functional calculus of
Ref.~\cite{Riec16a} shows that:
\begin{align}
\lim_{\omega \to 0} \text{Re} \StartMS (e^{i \omega} I - W)^{-1} \HWA = \EE - \frac{1}{2} \hmu ~.
\end{align}
This motivates introducing the \emph{excess-entropy spectrum} $\EEsp$:
\begin{align}
\label{eq:EESdef}
\EEsp
&\equiv \tfrac{1}{2} \bigl( \widetilde{\hsym}(\omega) - \Redundancy + \hmu \bigr) \\
&= \text{Re} \StartMS  (e^{i \omega} I - W)^{-1} \HWA + \tfrac{1}{2} \hmu
  ~.
\end{align}
The excess-entropy spectrum rather directly displays important frequencies of
apparent entropy reduction. For example, leaky period-$5$ processes have
a period-$5$ signature in the excess entropy spectrum.

As with its predecessors, the excess-entropy spectrum also has a natural
decomposition into two qualitatively distinct components:
\begin{align*}
\EEsp
  & = \mathcal{E}_{\rightsquigarrow}(\omega) + \mathcal{E}_{\multimap}(\omega)
  ~.
\end{align*}

The excess-entropy spectrum gives an intuitive and concise summary of the complexities associated with a process' predictability. For example, given a graph of the excess entropy spectrum, the past--future mutual information can be read off as the height of the continuous part of the function as it approaches zero frequency:
\begin{align*}
\EE & = \lim_{\omega \to 0} \EEsp \\
    & = \mathcal{E}_\text{c}(\omega=0)
	~.
\end{align*}
Indeed, the \emph{limit} of zero frequency is necessary due to the delta
function in the Fourier transform at exactly zero frequency:
\begin{align*}
\hmu & = \lim_{\epsilon \to 0}  \,
\frac{1}{\pi} \int_{-\epsilon}^{\epsilon} \EEsp \, d \omega
 ~.
\end{align*}
Reflecting on this, the delta function indicates one of the reasons the excess
entropy has been difficult to compute in the past. This also sheds light on the
role of the Drazin inverse: It removes the infinite asymptotic accumulation,
revealing the transient structure of entropy convergence.

We also have a spectral decomposition of the excess-entropy spectrum:
\begin{align*}
\EEsp & = \sum_{\lambda \in \Lambda_W} \sum_{m=0}^{\nu_\lambda - 1} 
  \text{Re} \biggl( \frac{\StartMS W_{\lambda, m} \HWA}{(e^{i \omega} -\lambda)^{m+1}}   \biggr) \\
   & = \sum_{m=0}^{\nu_0 - 1} \cos \bigl( (m+1) \, \omega \bigr) \StartMS W_0 W^m \HWA 
   \nonumber \\
   & \qquad
   + \sum_{\lambda \in \Lambda_W \setminus 0} \sum_{m=0}^{\nu_\lambda - 1} 
   \text{Re} \biggl( \frac{ \StartMS W_{\lambda, m} \HWA }{(e^{i \omega} -\lambda)^{m+1}} \biggr)
  ~,
\end{align*}
where, in the last equality, we assume that $W_0$ is real. This shows that, in
addition to the contribution of typical leaky modes of decay in entropy
convergence, the zero-eigenvalue modes contribute uniquely to the excess entropy
spectrum. In addition to Lorentzian-like spectral curves contributed by leaky
periodicities in the MSP, the excess-entropy spectrum also contains sums of
cosines up to a frequency controlled by index $\nu_0$, which corresponds to the
depth of the MSP's nondiagonalizability. This is simply the duration of
ephemeral synchronization in the time domain.

\subsection{Synchronization information} 

Once expressed in terms of the \syncMSP\ transition dynamic, the derivation of
the excess synchronization information $\SI'$ closely parallels that of the
excess entropy, only with a different ket $\ket{\cdot}$ appended. We calculate,
as before, finding:
\begin{align*}
\SI' & \equiv \sum_{L=0}^\infty \left[ \mathcal{H}(L) - \mathcal{H} \right] \\ 
  & = \sum_{L=0}^\infty
  \left[ \StartMS W^{L} \Hmxst - \StartMS W_1 \Hmxst \right] \\
  & = \sum_{L=0}^\infty  \StartMS
  \bigl[  (\underbrace{W - W_1}_{\equiv Q})^L - \delta_{L, 0} W_1 \bigr]
  \Hmxst \\
  & = - \underbrace{\StartMS W_1 \Hmxst}_{= \mathcal{H} }
  + \sum_{L=0}^\infty
  \underbrace{ \StartMS  Q^L }_{= \StartMS \mathcal{Q}^L}  \Hmxst \\
  & =  \StartMS
  \Bigl( \sum_{L=0}^\infty \mathcal{Q}^L \Bigr) \Hmxst - \mathcal{H} \\ 
  & =  \StartMS ( I - \mathcal{Q})^{-1} \Hmxst - \mathcal{H} \\ 
  & =  \StartMS ( I - Q)^{-1} \Hmxst - \mathcal{H}
  ~. 
\end{align*}
For an ergodic process where $W_1 = \ket{\one} \bra{\pi_W}$, this becomes:
\begin{align}
\SI' & = \StartMS \bigl( I - W + \ket{\one} \bra{\pi_W} \, \bigr)^{-1}
  \Hmxst - \mathcal{H}
  ~. 
\end{align}

From Eq.\ \eqref{eq: Drazin vs I less Q}, we see that the general solution for $\SI'$ takes on its most elegant form in terms of the Drazin inverse of $I-W$: 
\begin{align}
\SI' & =  \StartMS ( I - Q)^{-1} \Hmxst - \mathcal{H} \nonumber \\
& =  \StartMS ( I - Q)^{-1} \Hmxst - \StartMS W_1 \Hmxst \nonumber \\ 
& =  \StartMS \left[ ( I - Q)^{-1} - W_1 \right]  \Hmxst \nonumber \\ 
& = \StartMS  ( I - W)^{\mathcal{D}}  \Hmxst
  ~.
\label{eq: SI in terms of Drazin}
\end{align}

From Eq.~\eqref{eq: SI in terms of Drazin} and Eq.~\eqref{eq:ResolventAt1}, we
also see that the excess synchronization information has the general spectral
decomposition:
\begin{align}
\SI'
  & = \!\!\! \sum_{\lambda \in \Lambda_W \setminus \{ 1 \} } 
    \sum_{m=0}^{\nu_\lambda - 1} 
    \frac{1}{(1-\lambda)^{m+1}} 
    \StartMS  W_{\lambda, m}  \Hmxst
  ~. 
\end{align}

Again the form of Eq.~\eqref{eq: SI in terms of Drazin} suggests generalizing
synchronization information from a complexity measure to a complexity function
$\mathscr{S}(\omega)$. In this case, the result is simply related to the
Fourier transform of the two-sided myopic state-uncertainty $\mathcal{H}(L)$.

\subsection{Power spectra}

The extended complexity functions, $\EEsp$ and $\mathscr{S}(\omega)$ just introduced, give the same intuitive understanding for entropy reduction and synchronization respectively as the power spectrum $\Psp$ gives for pairwise correlation. Recall that the power spectrum can be written as:
\begin{align*}
P_\text{c}(\omega) & =  \bigl\langle \left| x \right|^2 \bigr\rangle + 
    2 \, \text{Re} \corrbra \left( e^{i \omega} I - \matHMM \right)^{-1} \corrket ~.
\end{align*}
We see that $\left( e^{i \omega} I - \matHMM \right)^{-1}$ is the resolvent of
$T$ evaluated along the unit circle $z = e^{i \omega}$ for $\omega \in [0, 2
\pi )$. Hence, by Part I's decomposition of the resolvent, 
the general spectral decomposition of the continuous part of the power spectrum
is:
\begin{align*}
P_\text{c}(\omega) & =  \bigl\langle \left| x \right|^2 \bigr\rangle + 
    2 
    \sum_{\lambda \in \Lambda_T} \sum_{m = 0}^{\nu_\lambda - 1} \, \text{Re}  \,
  \frac{ \corrbra T_{\lambda,m} \corrket  }{(e^{i \omega} - \lambda)^{m+1}} 
   ~.
\end{align*}
As with $\EEsp$ and $\mathscr{S}(\omega)$, all continuous 
frequency dependence of the power spectrum again lies simply and entirely
in the denominator of the above expression.

Analogous to Ref.~\cite{Riec14b}'s results, the power-spectrum delta
functions arise from the eigenvalues of $T$ that lie on the unit circle:
\begin{align*}
P_\text{d}(\omega) & =  
    \sum_{k = -\infty}^{\infty} 
  \sum_{\lambda \in \Lambda_T \atop |\lambda| = 1}
  2 \pi \, \delta( \omega - \omega_\lambda + 2 \pi k) 
  \nonumber \\
  & \qquad \qquad \qquad \times 
  \text{Re} \bigl( \lambda^{-1} \corrbra \, T_\lambda \corrket \bigr)
 ~,
\end{align*}
where $\omega_\lambda$ is related to $\lambda$ by $\lambda = e^{i
\omega_\lambda}$. An extension of the Perron--Frobenius theorem guarantees that
the eigenvalues of $T$ on the unit circle have index $\nu_\lambda = 1$.

Together, these equations yield structural constraints via particular
functional forms that are key to solving the inverse problem of inferring
process models from measured data.

\subsection{Almost diagonalizable dynamics}
\label{sec:AlmostDiag}

The nondiagonalizability that appears most commonly in prediction metadynamics
is of a special form that we call \emph{almost diagonalizable}: when all
eigenspaces except one---usually that associated with $\lambda=0$---are
diagonalizable subspaces. In the current setting, we say that a matrix is
\emph{almost diagonalizable} if all of its eigenvalues with magnitude greater
than zero have geometric multiplicity equal to their algebraic multiplicity. 

\begin{Def} 
$W$ is almost diagonalizable if and only if $g_\lambda = a_\lambda$ for all 
$\lambda \in \LWwoutZero \equiv \, \Lambda_W \setminus \{ 0 \}$. 
\end{Def}

Fortunately, we treat such nondiagonalizability straightforwardly using
$W^L$'s spectral decomposition for singular matrices. First off, Eq. \eqref{eq:
T^n generally} simplifies to:
\begin{align}
W^L & = 
  \sum_{\lambda \in \Lambda_W} \lambda^L  W_\lambda  
  + \sum_{m=1}^{\nu_0 - 1}  \delta_{L, m}  W_0 W^m 
  ~.    	
\label{eq: almost diagonalizable W^L spectral decomp}
\end{align} 

Then, to obtain the projection operators associated with each eigenvalue in
$\LWwoutZero$ for an almost diagonalizable matrix $W$, we use Part I's
expression for operators with index-one eigenvalues
with $\nu_\lambda = 1$ for all $\lambda \in \LWwoutZero$. Finding:
\begin{align} 
W_\lambda & = \left( \frac{W }{\lambda}  \right)^{\nu_0}
  \prod_{\zeta \in \LWwoutZero \atop \zeta \neq \lambda}
 	\frac{W - \zeta I }{\lambda - \zeta} 
  ~, 
\label{eq: proj operators for nonzero eigs of almost diag matrix}
\end{align}
for each $\lambda \in \LWwoutZero$. Or, when more convenient in a calculation,
we let $\nu_0 \to a_0 - g_0 + 1$ or even $\nu_0 \to a_0$ in Eq.\ \eqref{eq:
proj operators for nonzero eigs of almost diag matrix}, since multiplying
$W_\lambda$ by $W / \lambda$ has no effect.

With the set of projection operators $W_\lambda$ for all $\lambda \in
\LWwoutZero$ in hand, we can use the fact from Part I that projection operators
sum to the identity
to determine the projection operator associated with the zero eigenvalue: 
\begin{align*} 
W_0 = I - \sum_{\lambda \in \LWwoutZero} W_\lambda
  ~. 
\end{align*}
This is sometimes simpler and easier to automate than evaluating $W_0$ via the
methods of symbolic inversion and residues or via finding all left and right
eigenvectors and generalized eigenvectors.

Almost diagonalizable metadynamics play a prominent role in prediction for both
processes of finite Markov order and for the much more general class of
processes with broken partial symmetries that can be detected within a finite
observation window---the processes of finite symmetry-collapse discussed next.

\subsection{Markov order versus symmetry collapse} 

What if zero is the \emph{only} eigenvalue in the transient structure of a
process' MSP? That is, what if there are no loops in the \syncMSP\ transient
structure? The associated processes turn out to have finite Markov order.

For processes with finite Markov order $R$---such as, those whose support is a
\emph{subshift of finite type} \cite{Lind95a}---the entropy-rate approximates
not only converge but also become equal to the true entropy rate when conditioning on long enough histories.  Explicitly, for $\ell \geq R+1$ \cite{Crut01a}:
\begin{align}
\hmu(\ell) - \hmu = 0
  ~, 
\label{eq:FirstMarkovOrderDef}
\end{align}
or, equivalently, for $L \geq R$:
\begin{align*}
\StartMS W^L \HWA - \langle \pi_W \HWA = 0
  ~.
\end{align*}

For a finite-order Markov process, all MSP transient states must have
identically zero probability after $R$ time-steps. The only way to achieve this
is if the \syncMSP's transient structure is an acyclic directed graph with all
probability density flowing away from the unique start-state down to the
recurrent component. 
This means that all eigenvalues associated with the transient states are zero. Moreover, the index of the zero-eigenvalue of the \eM's \syncMSP\ is \emph{equal} to the Markov order for finite Markov-order processes. 
That is, if $\Lambda_W \setminus \Lambda_T = \{ 0 \}$, then:
\begin{align*}
\nu_0(W) = R
  ~.
\end{align*}

In contrast, for stochastic processes whose support is a \emph{strictly sofic
subshift} \cite{Lind95a}, the Markov order diverges, but $\nu_0$ can vanish or
be finite or infinite. Yet, in either the finite-type or sofic case, $\nu_0$
still tracks the duration of exact state-space collapse within the transient
dynamics of synchronization. This suggests that $\nu_0$ captures the
\emph{index of broken symmetries} for strictly sofic processes, in analogy to
the Markov order for subshifts of finite type. The name
\emph{symmetry-collapse} captures the essence of $\nu_0$'s role in both cases.

Let's explain. In the first $\nu_0$ time-steps, symmetries are broken that
synchronize an observer to the process. For the simple period-two process
$\dots 010101010 \dots$ the ``symmetry'' that is broken is the degeneracy of
possible phases---the $0$ phase or the $1$ phase of the period-$2$ oscillation.
Initially, without making a measurement the two phases are indistinguishable.
After a single observation, though, the observer learns the phase and is
completely synchronized to the process. Hence, $\nu_0 = 1$ for this order-1
Markov process. Simple periodic processes with larger periods have a longer
time before the phase information is fully known; hence, their larger Markov
order.

For the more complex strictly sofic processes, there may also be symmetries,
such as phase information, that are completely broken within a finite amount of
time. However, this is only part of the overall transient metadynamics of
synchronization. And so, the symmetries completely broken within the
symmetry-collapse epoch occur in addition to lingering state uncertainties
about a strictly sofic process. As a practical matter, a process'
predictability is often substantially enhanced through the finite epoch of
symmetry-collapse.  This becomes apparent in the examples to follow.

\begin{figure*}
\centering
\subfloat[How spectra emanate from eigenvalues: \emph{Coronal spectrogram}
	(far right) combines a process' eigenvalues $\Lambda_T$ (far left) of the
	hidden linear dynamic $T$ together with a frequency-dependent function
	$P(\omega)$ (middle) by wrapping the latter around the unit circle.
	]
	{\label{fig:CoronalSpectrogram}
\begin{overpic}[width=1.\textwidth,unit=2mm] 
      {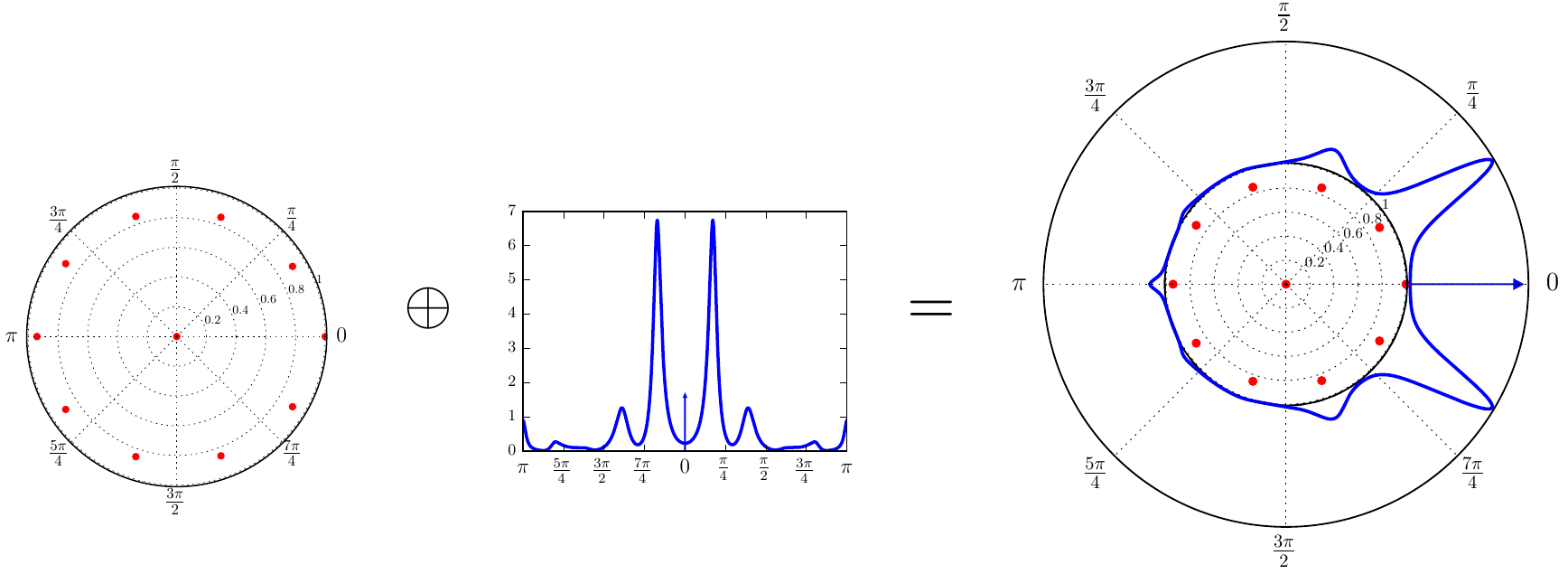}
    \put(20,12){\scalebox{0.7}{Re$(\lambda)$}}  
    \put(11,23){\scalebox{0.7}{Im$(\lambda)$}}  
    \put(52,26){\includegraphics[width=0.1\textwidth]{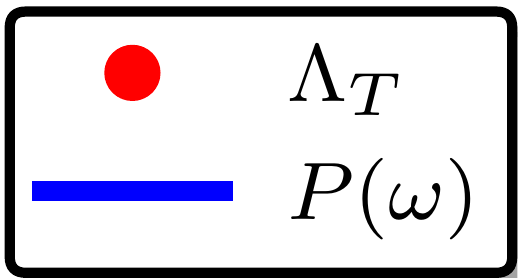}}
    \put(39,3.5){$\omega$}
    \put(37.5,22){$P(\omega)$}
    \put(9,25.7){$\Lambda_T$}
\end{overpic}
}
\\
\subfloat[\emph{Coronated horizon} (far right) combines the frequency-dependent
	function $f(\omega)$ (far left) together with a process' eigenvalues
	$\Lambda_A$ of the hidden linear dynamic by unwrapping the unit circle.]
	{\label{fig:CoronatedHorizon}
\begin{overpic}[width=1.\textwidth,unit=2mm] 
      {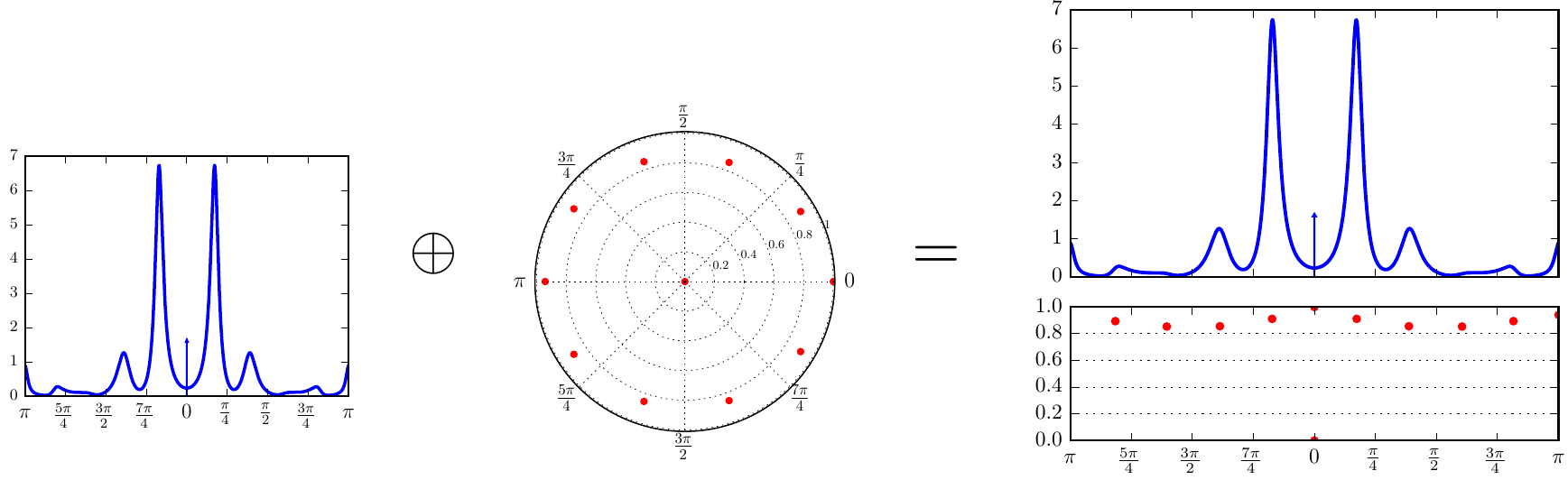}
    \put(79,21){\includegraphics[width=0.1\textwidth]{img/PS_legend.pdf}}
    \put(10.5,1.5){$\omega$}
    \put(9,20){$P(\omega)$}
    \put(38,23){$\Lambda_T$}
    \put(74.5,-0.5){$\omega$}
    \put(56.5,5){\rotatebox{90}{$|\lambda|$}}
\end{overpic}
}
\caption{Spectra and eigenvalues: (a) Coronal spectrogram and (b) coronated
	horizon.
	}
\label{fig:CoronalIntro}
\end{figure*}

\section{Spectral Analysis via Coronal Spectrograms}
\label{sec:CoronalSpec}

Coronal spectrograms are a broadly useful tool in visualizing complexity
spectra, from power spectra to excess entropy spectra. They were recently
introduced by Ref.~\cite{Riec14b} to demonstrate how diffraction patterns of
chaotic crystals emanate from the eigenvalue spectrum of the hidden spatial dynamic of stacked modular layers.

Coronal spectrograms display any frequency-dependent measure $f(\omega)$ of a
process wrapped around the unit circle while showing the eigenvalues
$\Lambda_T$ of the relevant linear dynamic $T$ within the unit circle in the
complex plane. Figure~\ref{fig:CoronalSpectrogram} gives an example. This is
appropriate for discrete-domain (e.g., discrete-time or discrete-space)
dynamics. For continuous-time dynamics, the coronal spectrogram unwraps into
what we call the \emph{coronated horizon}, via the familiar
discrete-to-continuous conformal mapping of the inside of the unit circle of
the complex plane to the left half of the complex plane~\cite{Latn98}.
Figure~\ref{fig:CoronatedHorizon} displays a discrete-time version of the
coronated horizon.  Ultimately, either the coronal spectrogram or coronated
horizon yield the same information and lend the same important lesson: the
eigenvalues of the hidden linear dynamic control allowed system behaviors.

\begin{figure*}
\subfloat[$(4\text{-}3)$-GM Process of the $(R\text{-}k)$-Golden Mean
	family with $0 \leq k = \nu_0(\zeta) \leq R = \nu_0(\mathcal{W}) < \infty$,
	which generates processes with finite but tunable Markov-order $R$ and
	cryptic-order $k$.]
	{\label{fig:43Rk}
\includegraphics[width=0.3\textwidth]{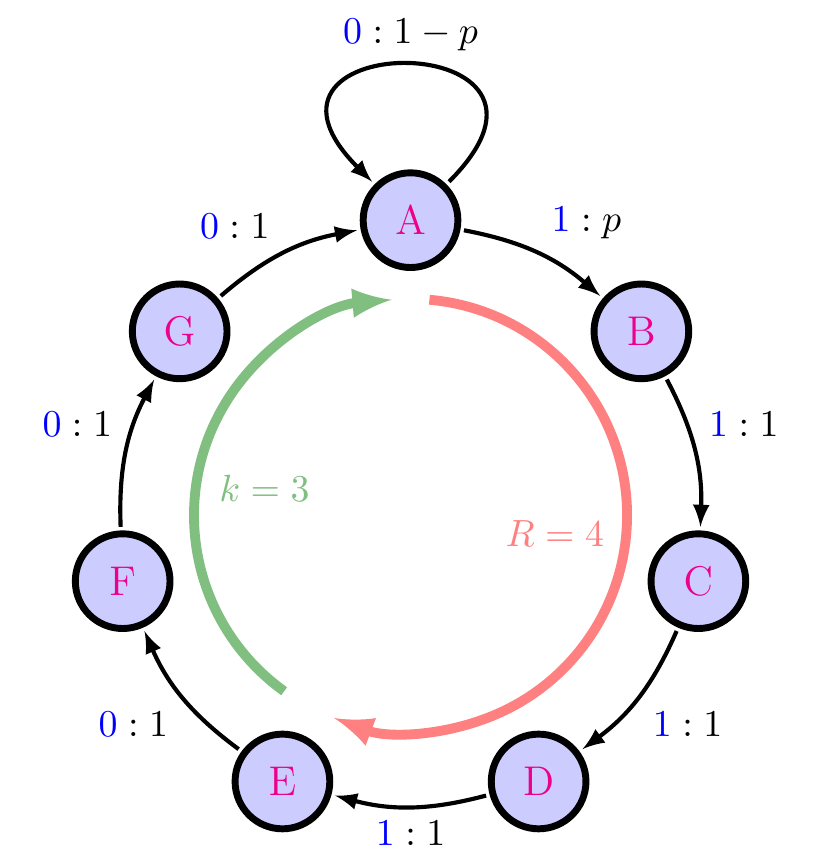}
}
\subfloat[$(4\text{-}3)$-GP-$(3)$ Process of the
	$(\nu_0(\mathcal{W})\text{-}k)$-Golden Parity-$(P)$ family with
	$0 \leq k = \nu_0(\zeta) \leq \nu_0(\mathcal{W})
	< R = \infty$ whenever $P>1$, generates processes with infinite
	Markov-order $R$, tunable finite cryptic-order $k$, and tunable finite
	symmetry-collapse index $\nu_0(\mathcal{W})$.]
	{\label{fig:43GP3}
\includegraphics[width=0.3\textwidth]{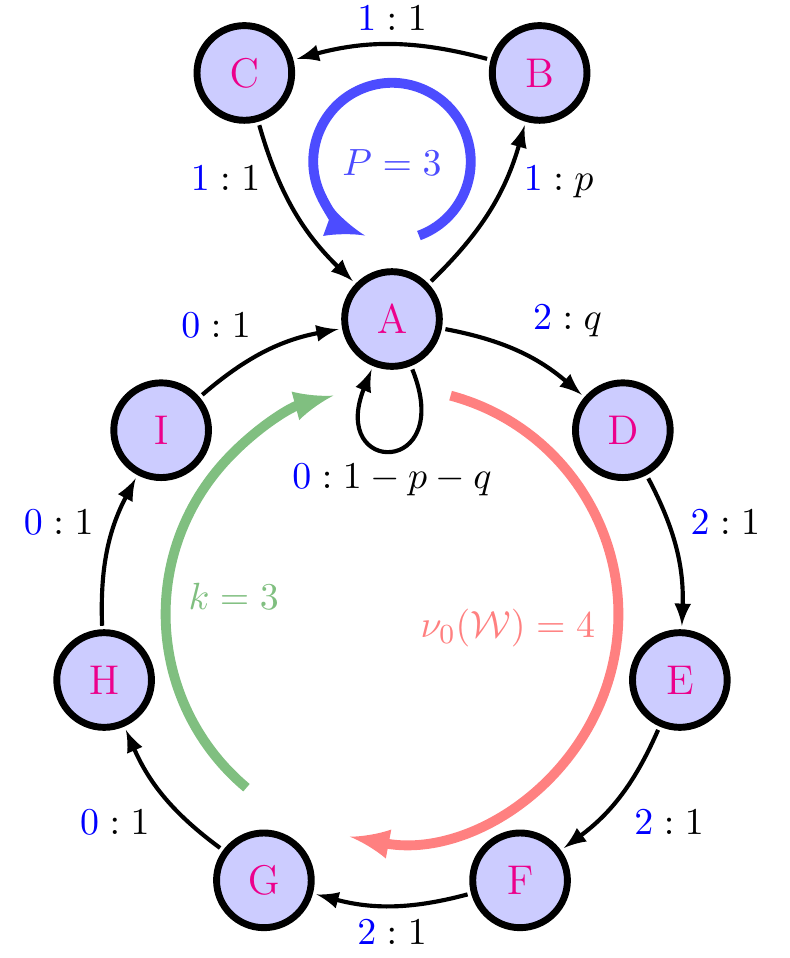}
}
\subfloat[$(4\text{-}3)$-GPZ-$(3\text{-}3)$ Process of the
	$(\nu_0(\mathcal{W})\text{-}\nu_0(\zeta))$-Golden Parity-$(P\text{-}Z)$
	family with $0 \leq \nu_0(\zeta) \leq \nu_0(\mathcal{W}) < k = R = \infty$
	whenever $Z > 1$. Markov order is infinite whenever either $P > 1$ or $Z >
	1$.  Cryptic-order is infinite when $Z > 1$. This family generates
	processes with finite but tunable symmetry-collapse index
	$\nu_0(\mathcal{W})$ and cryptic index $\nu_0(\zeta)$.]
	{\label{fig:43GPZ33}
\includegraphics[width=0.3\textwidth]{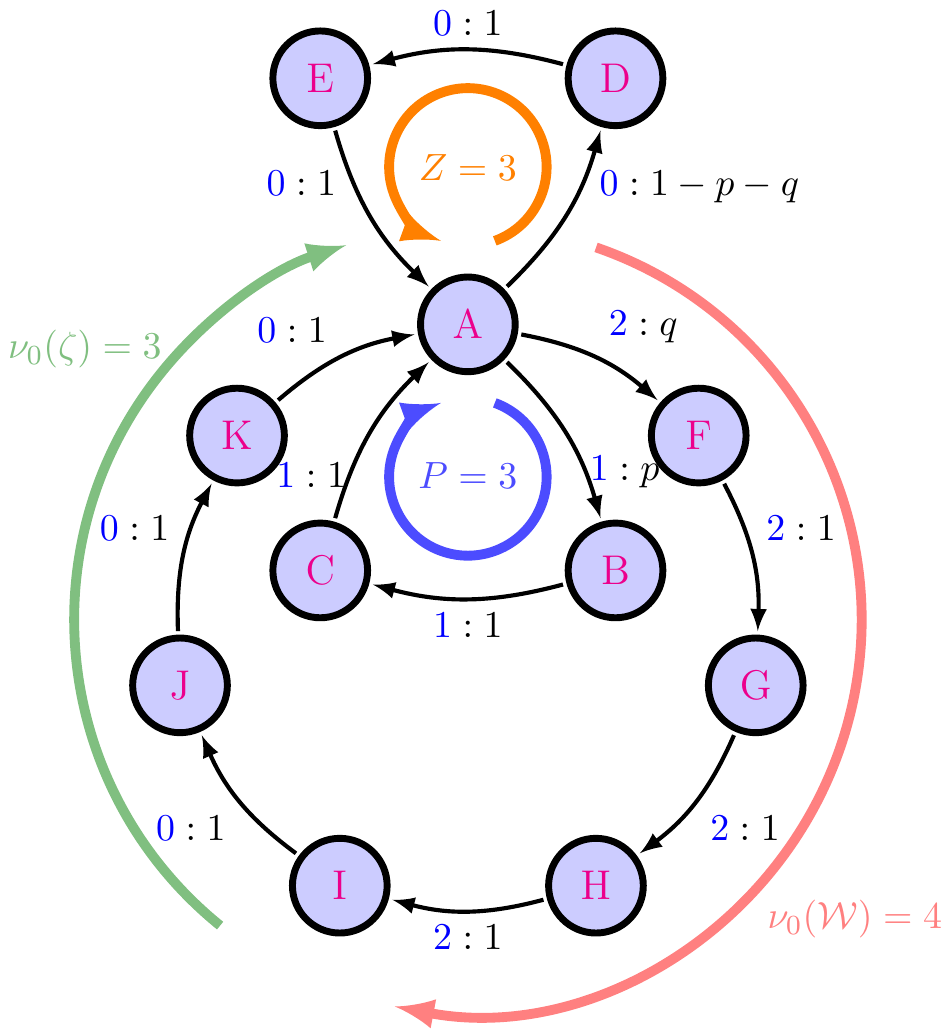}
}
\caption{Process families for exploring the roles of and interplay
	between Markov-order $R$, cryptic-order $k$, the symmetry-collapse index
	$\nu_0(\mathcal{W})$ of the zero eigenvalue of the synchronizing dynamic
	over mixed states, and the cryptic index $\nu_0(\zeta)$ of the zero
	eigenvalue of the cryptic operator presentation. We always have $k \leq R$
	and $\nu_0(\zeta) \leq \nu_0(\mathcal{W})$. Whenever $\Lambda_{\mathcal{W}}
	= \Lambda_{T} \cup \{ 0 \}$, $R$ is finite, $R = \nu_0(\mathcal{W})$ and $k
	= \nu_0(\zeta)$. Whenever $\Lambda_{\zeta} = \Lambda_{T} \cup \{ 0 \}$,
	$k$ is finite, whether or not $R$ is, and $k = \nu_0(\zeta)$. When $k$ or
	$R$ is infinite, the cryptic index and symmetry-collapse index reveal more
	nuanced features of the cryptic and synchronization dynamics.
}
\end{figure*}

Coronal spectrograms demonstrate that complex systems behave according to the
spectrum of their hidden linear dynamic. The relevant frequency-dependent
measure $f(\omega)$ emanates from the nonzero eigenvalues of the hidden linear
dynamic: the closer eigenvalues approach the unit circle, the sharper the
observed peaks. At one extreme, one observes Bragg-like reflections
(delta-function contributions) when the eigenvalues fall on the unit circle.
The collection of diffuse peaks observed is a sum of Lorentzian-like and, what
we might call, super-Lorentzian-like line profiles. Indeed, the Lorentzian-like
line profiles are the discrete-time version of a Lorentzian curve. While the
continuous-domain Lorentzian is given by $\text{Re}\bigl( \frac{c}{\omega -
\lambda} \bigr)$, the continuous-to-discrete conformal mapping $\omega \to e^{i
\omega}$ directly yields our discrete-domain analog $\text{Re} \bigl(
\frac{c}{e^{i \omega} - \lambda} \bigr)$. The super-Lorentzian-like line
profiles have the form $\text{Re}\bigl[ ( \frac{c}{e^{i \omega} - \lambda})^{n}
\bigr]$.

Zero eigenvalues also contribute to $f(\omega)$, but only sinusoidal
contributions of discrete increments from $\cos(\omega)$ up to $\cos(\nu_0 \,
\omega)$. Since these are qualitatively distinct from the super-Lorentzian
contributions and do not emanate radially from the eigenvalues the same way
contributions from nonzero eigenvalues do, coronal spectrograms are most useful
for understanding the contributions of nonzero eigenvalues. Nevertheless, the
two contributions can be usefully disentangled, as shown later.

We use both coronal spectrograms and coronated horizons to visualize various
features in the examples to follow.

\section{Examples}
\label{sec:Examples}

\subsection{Golden Mean Processes}

To explore finite Markov order in relation to various complexity measures let's
consider the $(R\text{-}k)$-Golden Mean (GM) Processes~\footnote{Historically,
the name ``Golden Mean'' derives from the $(1\text{-}1)$-Golden Mean Process
that forbids two consecutive ${\color{blue}1}$s. It describes the symbolic
dynamics of the chaotic one-dimensional shift map over the unit interval with
generating partition, when the slope of the map is the golden mean $\phi = 1 +
1 / \phi = (1 + \sqrt{5})/2$~{\cite{Will04}}. Moreover, consistency with the
stationary stochastic process generated implies $1-p = 1/\phi$. Generalizing to
arbitrary $p$ and especially to arbitrary $R$ and $k$ takes us away from the
eponymous property into a setting sufficiently general to study the generic
behavior of order-$R$ Markov processes and the complexities of predicting
them.}. This process family describes a unique transition-parametrized process
for each Markov-order $R \in \bigl\{ \nu \in \mathbb{Z} : \nu \geq 1 \bigr\}$
and each cryptic-order $k \in \bigl\{ \kappa \in \mathbb{Z} : 1 \leq \kappa
\leq R \bigr\}$. The \eM\ for the $(4\text{-}3)$-Golden Mean Process is shown
in Fig.~\ref{fig:43Rk}. From this the construction of all other
$(R\text{-}k)$-Golden Mean processes can be discerned. In words,
$(R\text{-}k)$-Golden Mean Processes are binary with alphabet $\Abet = \{
{\color{blue}0}, {\color{blue}1} \}$ and if the most recent history consists of
at least $k$ consecutive ${\color{blue}0}$s (and no ${\color{blue}1}$s since
then) then there is a probability $p$ of next observing a ${\color{blue}1}$ and
a probability $1-p$ of simply seeing another ${\color{blue}0}$. This entails
$R$ consecutive ${\color{blue}1}$s followed by at least $k$ consecutive
${\color{blue}0}$s.

\begin{figure}
\includegraphics[width=0.4\textwidth]{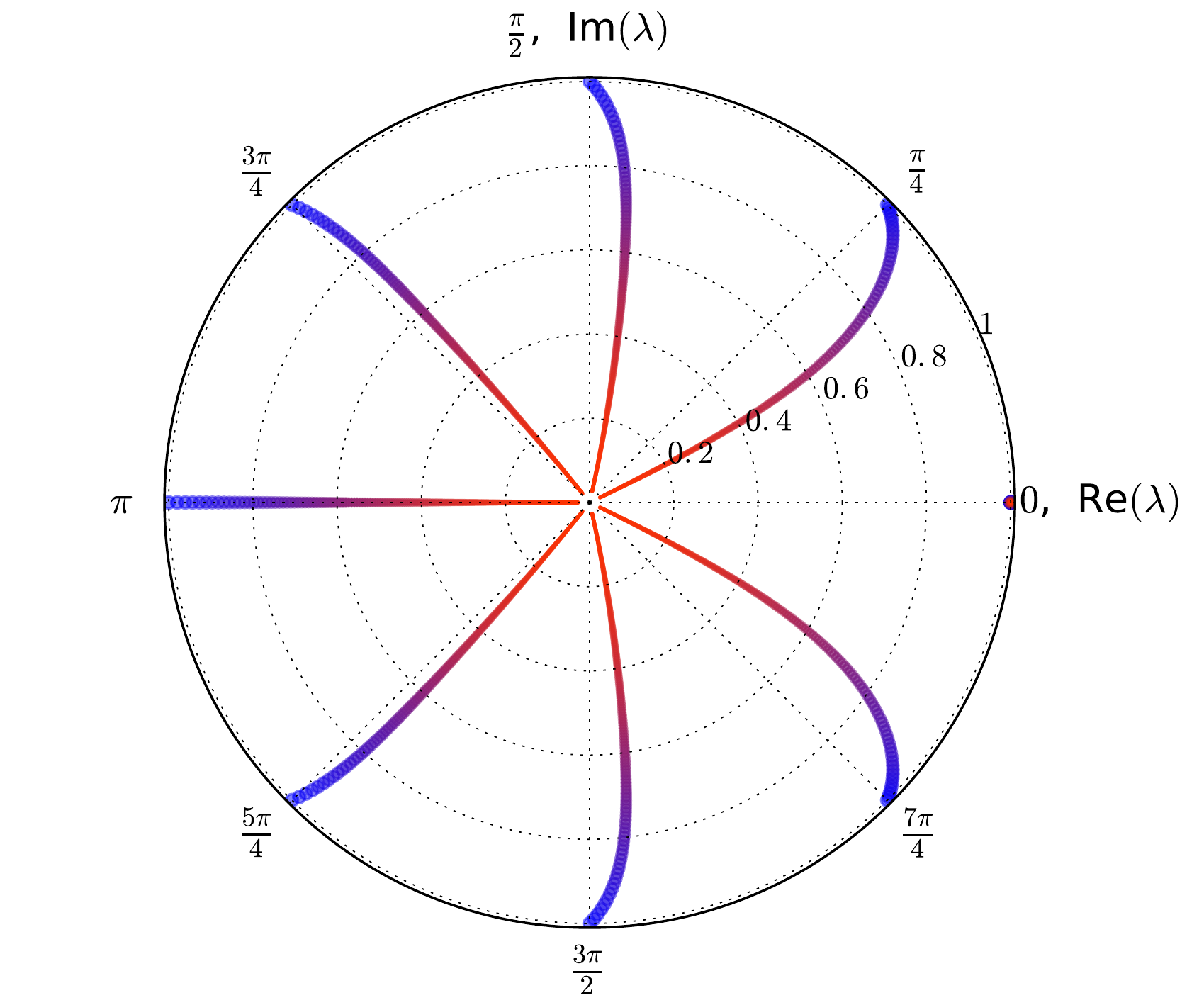}
\caption{Evolution of eigenvalues $\Lambda_T$ of the recurrent component of the
	(5--3)-GM Process's \eM. Displayed within the unit circle of the complex
	plane, the trajectory of each eigenvalue follows a line that starts thick
	blue and ends thin red as the transition parameter $p$ evolves from $1$ to
	$0$. In addition to the seven eigenvalues that move from the nontrivial
	eighth roots of unity towards zero along nonlinear trajectories, the
	eigenvalue at $z=1$ does not change with $p$.
	}
\label{fig:53GM_TeigsEvolution}
\end{figure}

The eigenvalues of the internal state-to-state transition matrix of the
\eM's recurrent component are: 
\begin{align*}
\Lambda_T = \Bigl\{ \lambda \in \mathbb{C}: \bigl( \lambda - (1-p) \bigr) \lambda^{R+k-1} = p \Bigr\}
~.
\end{align*}
In the limit of $p \to 1$, all $(R\text{-}k)$-Golden Mean Processes become
perfectly periodic. In this limit, the eigenvalues are evenly distributed on
the unit circle:
\begin{align*}
\Lambda_T \to \bigl\{ e^{i n 2 \pi / (R + k)} \bigr\}_{n=0}^{R+k-1}
  ~.
\end{align*}
At the other extreme, as $p \to 0$, all eigenvalues
evolve to zero, except the stationary eigenvalue at $z=1$. At any setting of
$p$, the nonunity eigenvalues lie approximately on a circle within the complex
plane whose radius decreases nonlinearly from $1$ to $0$ as $p$ is swept from
$1$ to $0$. Simultaneously, this circle's center moves from the origin to a
positive real value and back to the origin as $p$ is swept from $1$ to $0$.
Figure~\ref{fig:53GM_TeigsEvolution} shows how the eigenvalues of the
$(5\text{-}3)$-Golden Mean Process evolve over the full range of $p$ as it
sweeps from $1$ to $0$.

In contrast to the $p$-dependent spectrum of the recurrent structure just
discussed, the only eigenvalue corresponding to the transient structure of the
\syncMSP\ is equal to zero, regardless of the transition parameter $p$.  Recall
that this is necessarily true for any process with finite Markov order. Hence,
$\Lambda_{\mathcal{W}} = \Lambda_T \cup \{ 0 \}$, with $\nu_0(\mathcal{W}) =
R$.  The cryptic structure is similar: $\Lambda_{\zeta} = \Lambda_T \cup \{ 0
\}$, with $\nu_0(\zeta) = k$, where $\zeta$ is the state-to-state transition
matrix of the cryptic operator presentation.

Table~\ref{table:FiniteR} compares the \eMs, autocorrelation, power spectra,
MSPs, myopic entropy rates, and myopic state uncertainties for three
$p$-parametrized examples of $(R\text{-}k)$-GM processes.

The autocorrelation of each process captures their `leaky periodic' behaviors:
The leakiness originates from the self-transition at state ${\color{magenta}A}$
that adds a phase-slip noise to otherwise $(R+k)$--periodic behavior.
Moreover, each process' phase, and so its \eM's internal state, is uniquely
identified after $R$ observations. This corresponds to the depth of the
\syncMSP\ tree-like structure $\nu_0(\mathcal{W}) = R$, the convergence of the
myopic entropy rate $\hmu(L)$ to the true entropy rate $\hmu$ when conditioning
on observations of finite block-length $L-1=R$, and the complete loss of causal
state uncertainty $\mathcal{H}(L)$ after $L=R$ observations.

\begin{table*}
\begin{center}
\includegraphics[width=0.8\textwidth]{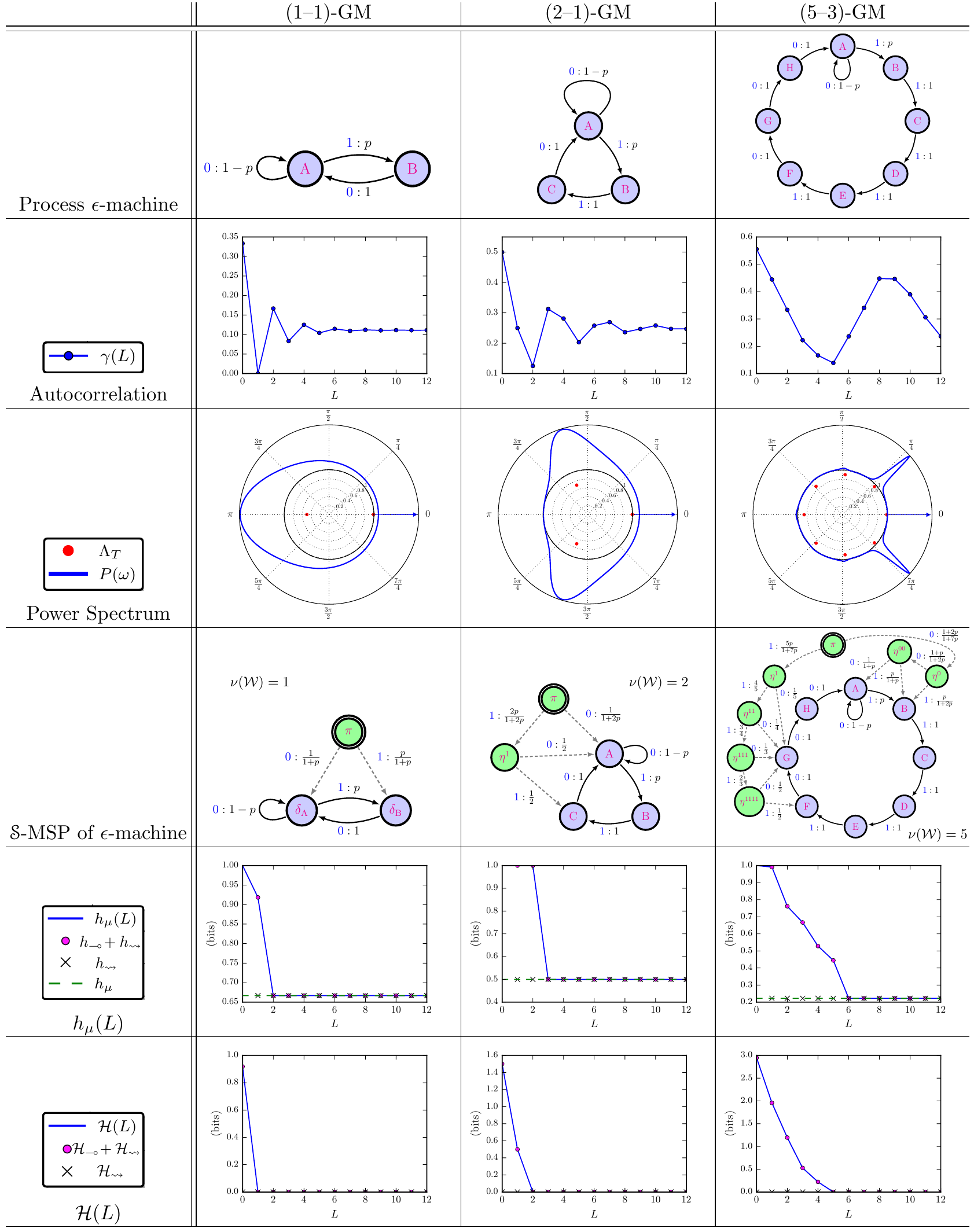}
\end{center}
\caption{Complexity analyses for finite Markov-order processes. Quantitative
	data used $p = 1/2$.
	}
\label{table:FiniteR}
\end{table*}

A paradigm of finite Markov order, the $(5\text{-}3)$-Golden Mean Process has a
strictly tree-like structure in its MSP's transients, which have a maximum
depth equal to both $\nu(\mathcal{W})$ and its Markov order of $5$.

These analyses illustrate the typical behaviors of complexity measures for
finite Markov-order processes. We next investigate examples of infinite
Markov-order processes to draw attention to the characteristic differences of
nonzero eigenvalues in their MSP transient structures.

\subsection{Even Process}

The Even Process, shown in the first column of Table~\ref{table:infiniteR}, is
a well known example of a stochastic process that cannot be generated by any
finite Markov-order approximation, yet it is described by a simple two-state
HMM.

Infinite Markov order, in this case, stems from the fact that the process
generates only an even number of consecutive $1$s, between $0$s. The countably
infinite set of Markov chain states necessary to track this parity reflects the
infinite order. Moreover, the surplus entropy rate $\hmu(L) - \hmu$ incurred
when using a finite order-$(L-1)$ Markov approximation vanishes only
asymptotically, being the sum of decaying exponentials. (See
Table~\ref{table:infiniteR}.) This is in stark contrast to the myopic entropy
rate for the finite Markov order processes of Table~\ref{table:FiniteR}. For
them $\hmu(L)$ drops to $\hmu$ \emph{exactly} at $L = R+1$. Similarly, the
average state uncertainty $\mathcal{H}(L)$ for infinite Markov processes
converges only asymptotically---and with the same set of decay rates as
$\hmu(L)$---to its asymptotic value of $0$. (This curve is not shown in
Table~\ref{table:infiniteR} for lack of space.) Such long-lived decay is driven
by nonzero eigenvalues in the \syncMSP\ transient structure.

The Even Process is a relatively simple example of an infinite Markov-order
process. As expected for infinite Markov-order, its MSP's transient structure
had nonzero eigenvalues. Generally, though, two ranges of contribution are to
be expected in synchronization dynamics. The first is a finite-horizon
contribution to the past-future mutual information, corresponding to completely
ephemeral zero eigenvalues in the MSP's transient structure. The second is an
infinite-horizon contribution to the past--future mutual information, arising
from nonzero eigencontributions.

\begin{table*}
\begin{center}
\includegraphics[width=0.8\textwidth]{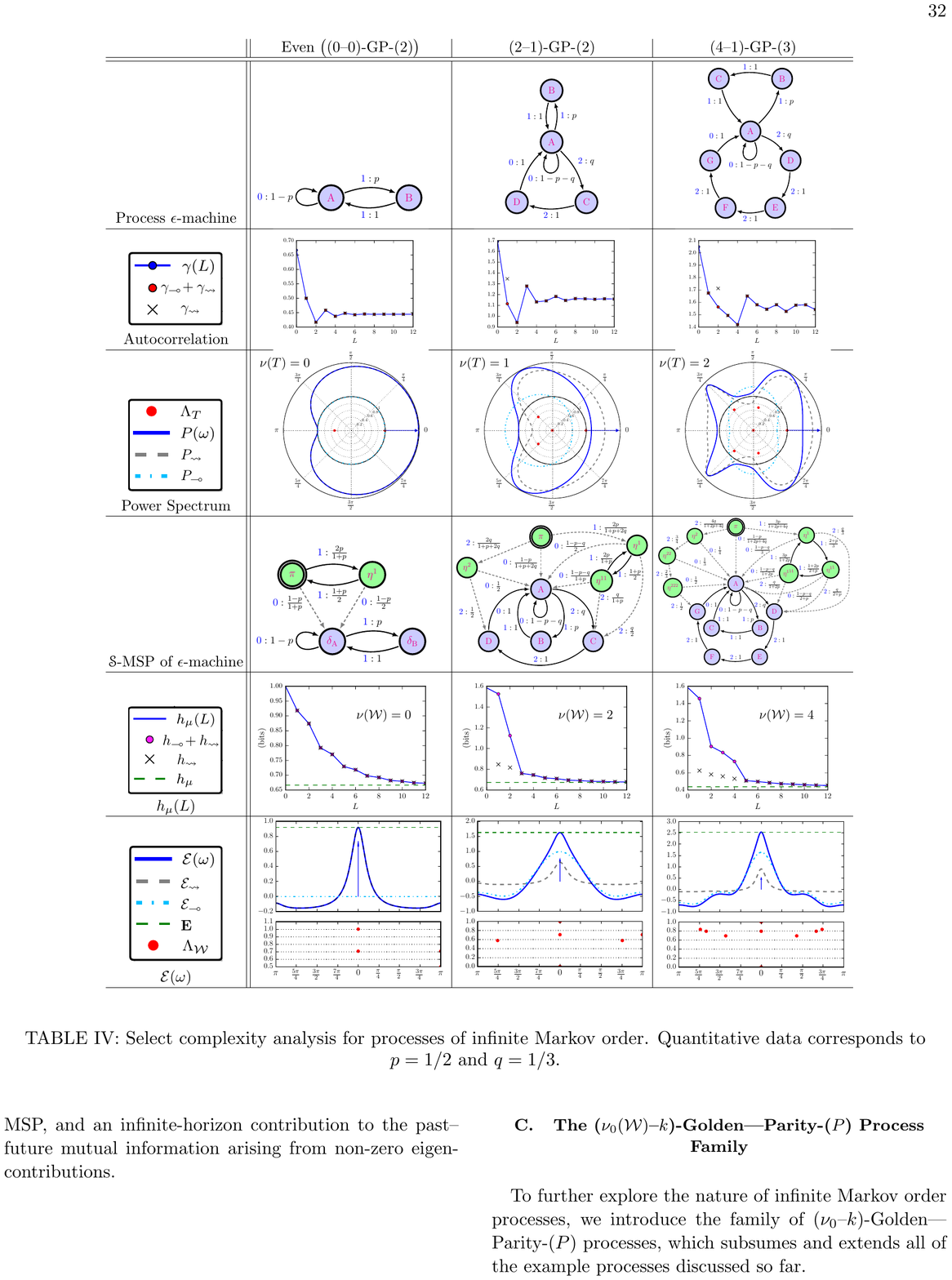}
\end{center}
\caption{Complexity analyses for infinite Markov-order processes. Quantitative
	data used $p = 1/2$ and $q = 1/3$.
	}
\label{table:infiniteR}
\end{table*}

\subsection{Golden--Parity Process Family}

To further explore the nature of infinite Markov order processes, we introduce
the $(\nu_0\text{-}k)$-Golden-Parity-$(P)$ Processes. This family subsumes and
extends the examples analyzed so far. The role of each parameter is explained
in Fig.~\ref{fig:43GP3}, which displays a state-transition diagram of the
$(4\text{-}3)$-GP-$(3)$ Process' \eM.
 
If $P=1$, the family reduces to the $(\nu_0\text{-}k)$-Golden Mean Process
family, with tunable Markov $R=\nu_0(W)$ and cryptic $k$ orders. That is,
$(\nu_0(\mathcal{W})\text{-}k)$-GP-$(1)$ = $(\nu_0(\mathcal{W})\text{-}k)$-GM.
However, the Markov order becomes infinite whenever $P > 1$. In this case the
index $\nu_0(\mathcal{W})$ of the \syncMSP's zero-eigenvalue---which controls
the finite duration necessary to resolve all broken symmetries---and the
cryptic order $k$ can still be tuned independently. The Even Process considered
earlier is the $(0\text{-}0)$-Golden-Parity-$(2)$ Process.

Three examples of $(\nu_0(\mathcal{W})\text{-}k)$-Golden-Parity-$(P)$ processes
are analyzed in Table~\ref{table:infiniteR}. The \syncMSP\ transient structure
for the second two clarifies the difference between (i) the symmetry collapse
associated with completely ephemeral transient states that are fully depleted
of probability density after $\nu_0(\mathcal{W})$ time-steps and (ii) the
long-lived leaky transients whose probability density only vanishes as
more-refined ambiguity is resolved.

Examining the myopic entropy convergence $\hmu(L)$, the effect of these
distinct routes to synchronization on the predictability can be seen: The
process is much more predictable, on average, after $\nu_0(\mathcal{W})$
time-steps. However, the average predictability of an infinite-Markov-order
process continues to increase with increasing observation window, albeit with
exponentially diminishing returns. In general, we showed that this asymptotic
convergence occurs as a sum of decaying exponentials from diagonalizable
subspaces and as the product of polynomials and exponentials in the case of
nondiagonalizable structures associated with nonzero eigenvalues. The apparent
oscillations under the exponential decays are completely described by the leaky
periodicities of the eigenvalues in the transient belief states.

Finally, note that the excess entropy spectrum $\mathcal{E}(\omega)$ shows the
frequency domain view of observation-induced predictability. $\EE =
\lim_{\omega \to 0} \mathcal{E}(\omega)$ is the total past--future mutual
information, which is also the excess entropy observed before full
synchronization. The $\nu_0(\mathcal{W})$ symmetry collapse contributes
significantly and early to the total excess entropy of the last two examples.
Whereas, the asymptotic tails of synchronization associated with leaky
periodicity of particular transient states of uncertainty accumulate their
contribution to excess entropy rather slowly.

In addition to new intuitions about convergence behaviors in stochastic
processes, the general and broadly applicable theoretical results here allow
novel numerical investigations and unprecedentedly-accurate analyses of
infinite-Markov-order processes. As an example of the latter, let us summarize
several of the exact results derived in App.~\ref{sec:ExactCalculation} for the
$(p, q)$-parametrized $(2\text{-}1)$-GP-$(2)$ process explored in
Table~\ref{table:infiniteR}'s second column.

Depending on whether the transition parameter $p$ is larger or smaller than $2
\sqrt{q} - q$, App.~\ref{sec:ExactCalculation} found qualitatively distinct
behaviors dominate the $(2\text{-}1)$-GP-$(2)$ process. This hints at a
general principle: behaviorally distinct regions are separated by a critical line in the $(p,q)$-parameter space along which the transition dynamic $T$ becomes nondiagonalizable. For $p < 2 \sqrt{q} - q$, the autocorrelation for  $|L| \geq 2$ has the exact solution:
\begin{align}
\gamma(L) & = \beta^2 +
\beta \, q^{|L| / 2} \, \text{Re} \bigl( \zeta \, e^{i \omega_\xi |L| } \bigr)
 ~,
\end{align}
where $\beta \equiv 2(p + 2q) / (1+p+2q)$, 
$\zeta \equiv (\xi +1)^2 (p \xi + 2 q) / (\xi ( \xi^3 + p \xi + 2 q ) )$,
$\xi \equiv -\frac{1}{2} (p+q) + i \frac{1}{2} \sqrt{4q - (p+q)^2}$, and
$\omega_\xi \equiv \tfrac{\pi}{2} + \arctan \bigl( (p+q)/\sqrt{4q - (p+q)^2} \bigr)$.
The corresponding power spectrum is:
\begin{align}
P(\omega) 
  & =  \frac{8q}{1+p+2q} +  \frac{2p}{1+p+2q} \bigl[ 1 - \cos(\omega) \bigr]
  \nonumber \\
  & \qquad + \beta \, \text{Re} \Bigl( \frac{\zeta \xi}{e^{i \omega} - \xi }
  + \frac{\zeta \xi}{e^{- i \omega} - \xi }  \Bigr) \nonumber \\
  & \qquad + 2 \pi \beta^2
  \sum_{k = -\infty}^{\infty} \delta( \omega + 2 \pi k)
  ~.
\end{align}

For any parameter setting, the metadynamic of observation-induced
synchronization to the $(2\text{-}1)$-GP-$(2)$ process is nondiagonalizable due
to the index-$2$ zero eigenvalue. This leads to a completely ephemeral
contribution to $\hmu(L)$ up to $L=2$. For $L \geq 3$, we find the myopic
entropy rate relaxes asymptotically to the true entropy rate according to: 
\begin{align*}
\hmu(L) - \hmu & =
\begin{cases}
\frac{-p \log p + (1+p) \log(1+p) - 2p}{\sqrt{p} (1+p+2q)} \, p^{L/2}
   & \text{for odd } L, \\
\frac{p \log p - (1+p) \log(1+p) + 2}{ (1+p+2q)} \, p^{L/2}
  & \text{for even } L,
\end{cases}
\end{align*}
where the process' true entropy rate is:
\begin{align*}
\hmu & = \frac{-q \log q - p \log p - (1-p-q) \log (1-p-q) }{1+p+2q}
  ~.
\end{align*}

Interestingly, while the autocorrelation at separation $L$ scales as $\sim
q^{L/2}$, the predictability of single-symbol transitions between slightly shifted histories of
length $L$ converges as $\sim p^{L/2}$---indicating two rather independent decay rates.

The amount of the future that can be predicted from the past is the total
mutual information between the observable past and observable future:
\begin{align*}
\EE & = 
\tfrac{(1-p-q) \log(1-p-q) - p \log p  - q \log q  - (1-p) \log( 1-p )  }{1+p+2q}
  \nonumber \\
  & \qquad + \log(1+p+2q)
  ~.
\end{align*}
However, to actually \emph{perform} prediction requires more memory than this
amount of shared information. Calculation of additional measures and more
detail can be found in App.~\ref{sec:ExactCalculation}.

To explore the structure in infinite-cryptic order processes, one can use the
more generalized family of
$(\nu_0(\mathcal{W})\text{-}\nu_0(\zeta))$-GP-$(P\text{-}Z)$ Processes. For
them, $\nu_0(\zeta)$ is the index of the zero-eigenvalue of the cryptic
operator presentation and the process has infinite cryptic order whenever
$Z>1$. Above, $Z=1$ and
$(\nu_0(\mathcal{W})\text{-}\nu_0(\zeta))$-GP-$(P\text{-}1)$ =
$(\nu_0(\mathcal{W})\text{-}\nu_0(\zeta))$-GP-$(P)$. Since the preceding
examples served well enough to illustrate the power of spectral decomposition,
our main goal, we leave a full analysis of this family to interested others.

\section{Predicting Superpairwise Structure}
\label{sec:RRX}

\begin{figure}
\includegraphics[width=0.3\textwidth]{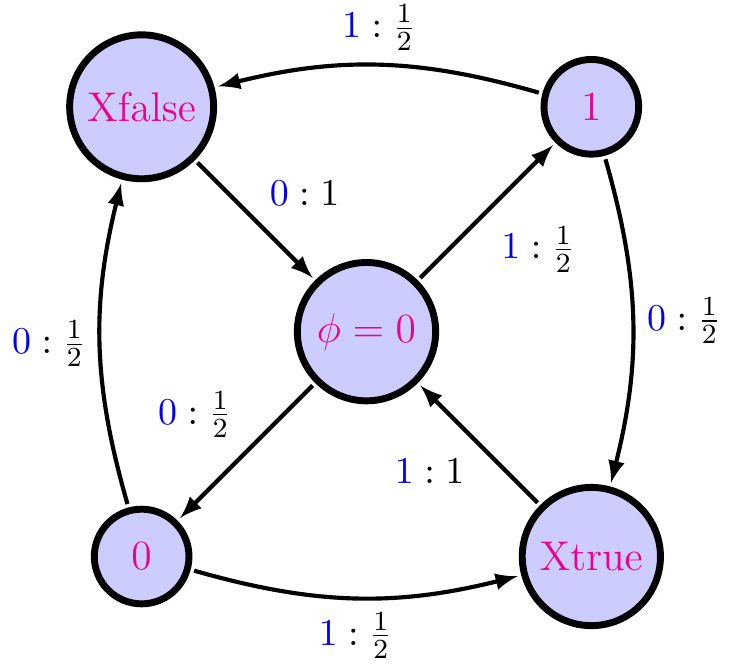}
\caption{RRXOR Process \eM.}
\label{fig:RRXeM}
\end{figure}

The Random--Random--XOR (RRXOR) Process is generated by a simple HMM.
Figure~\ref{fig:RRXeM} displays its five-state \eM. However, it illustrates
nontrivial, counterintuitive features typical of stochastic dynamic information
processing systems. The process is defined over three steps that repeat: (i) a
$\color{blue}0$ or $\color{blue}1$ is output with equal probability, (ii)
another $\color{blue}0$ or $\color{blue}1$ is output with equal probability,
and then (iii) the eXclusive-OR operation (XOR) of the last two outputs is output.

\begin{figure}
\includegraphics[width=0.4\textwidth]{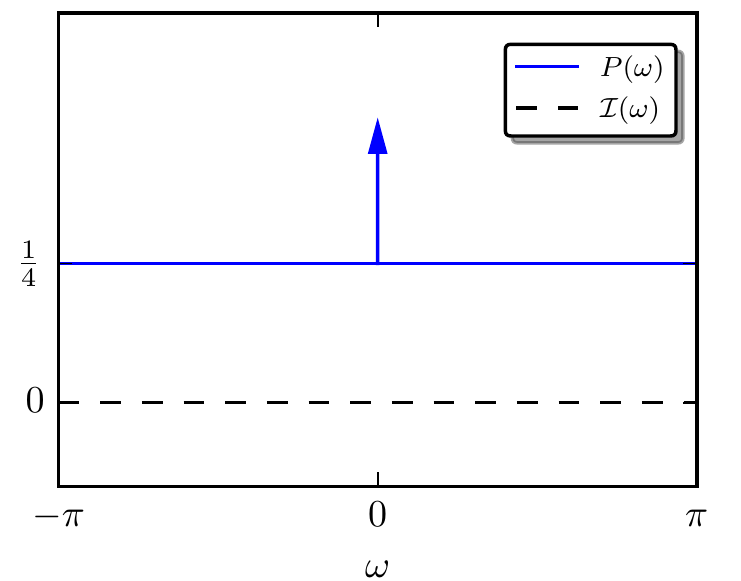}
\caption{Power spectrum $P(\omega)$ and POPI spectrum $\mathcal{I}(\omega)$ of
	the RRXOR Process: The first is flat and the second identically zero. One
	might incorrectly conclude the RRXOR Process is structureless white noise.
	}
\label{fig:RRXPowerSpectrum}
\end{figure}

Surprisingly, but calculations easily verify, there are no pairwise
correlations. All of its correlations are higher than second order. One
consequence is that its power spectrum is completely flat---the signature of
white noise; see Fig.~\ref{fig:RRXPowerSpectrum}. This would lead a casual
observer to incorrectly conclude that the generated time series has no
structure. In fact, a white noise spectrum is an indication that, if
structure it present, it must be hidden in higher-order correlations.

The RRXOR Process clearly is not structureless---via the exclusive OR, it
transforms information in a substantial way. We show that the complexity
measures introduced above can detect this higher-order structure. However, let
us first briefly consider \emph{why} the correlation-based measures fail to
detect structure in the RRXOR Process.

It is sometimes noted that information measures are superior to standard
measures of correlation since they capture nonlinear dependencies, while the
standard correlation relies on linear models. And so, we can avoid this problem
by using the \emph{information correlation} $\I[X_0; X_\tau]$ rather than
autocorrelation. Analogous to autocorrelation, it too has a spectral
version---the \emph{power-of-pairwise information (POPI) spectrum}:
\begin{align}
\mathcal{I}(\omega) 
  \equiv - \H(X_0) + \lim_{N \to \infty} \sum_{\tau = -N}^{N} 
  e^{- i \omega \tau} \I[X_0; X_\tau]
  ~.
\end{align}
It is easy to show that $\mathcal{I}(\omega) = 0$ for the RRXOR Process.
Hence, as Fig.~\ref{fig:RRXPowerSpectrum} showed, such measures are not
sufficient to detect even simple computational structure, since they only can
detect \emph{pairwise} statistical dependencies.

\begin{figure}
\begin{center}
\begin{overpic}[width=0.46\textwidth,unit=1mm] 
      {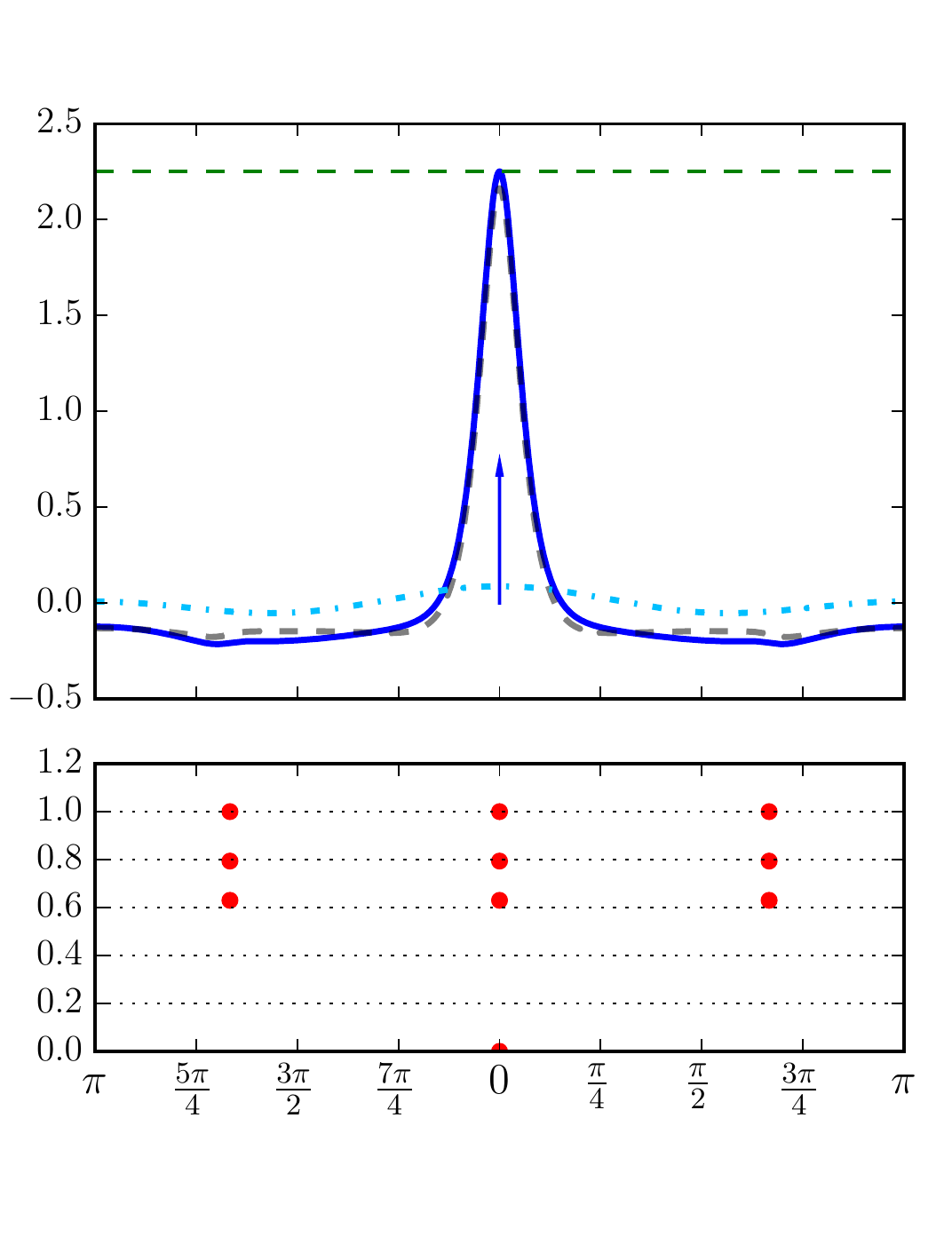}
  \put(58,60){\includegraphics[width=0.1\textwidth]{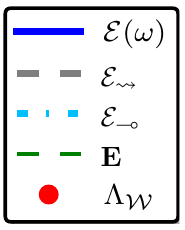}}
  \put(-1,20){\rotatebox{90}{$|\lambda|$}}
  \put(42.5,1){\scalebox{1.1}{$\omega$}}
\end{overpic}
\end{center}
\caption{Excess entropy spectrum of the RRXOR Process, together with the
	eigenvalues of the \syncMSP\ transition matrix $\mathcal{W}$. Among the
	power spectrum, POPI spectrum, and excess entropy spectrum, only the excess
	entropy spectrum is able to detect structure in the RRXOR Process since the
	structure is beyond pairwise. The eigenspectrum of the MSP of the RRXOR
	\eM\ and the excess entropy spectrum both indicate that the RRXOR Process
	is indeed structured, with both ephemeral symmetry breaking and leaky
	periodicities in the convergence to optimal predictability.
	}
\label{fig:RRX_EE_spectrum}
\end{figure}

In stark contrast, the excess entropy spectrum $\mathcal{E}(\omega)$ does
identify the structure of hidden dependencies in the RRXOR Process; see
Fig.~\ref{fig:RRX_EE_spectrum}. Why? The brief detour through power spectra,
information correlation, and POPI spectra brings us to a deeper understanding
of why $\mathcal{E}(\omega)$ is successful at detecting nuanced computational
structure in a time series. Since it partitions all random variables throughout
time, the excess entropy itself picks up any systematic influence the past has
on the future. The excess entropy \emph{spectrum} further identifies the
frequency decomposition of any such linear or nonlinear dependencies. In short,
all multivariate dependencies contribute to the excess entropy spectrum.

Let us now consider the hidden structure of the RRXOR Process in more detail.
With reference to Fig.~\ref{fig:RRXeM}, we observe that the expected
probability density over causal states evolves through the \eM\ with a
period-$3$ modulation. In a given realization, the particular symbols emitted
after each phase resetting ($\phi = 0$) break symmetries with respect to
which ``wings'' of the \eM\ structure are traversed. This is reflected in $T$'s
eigenvalues: the three roots of unity $\{ e^{i n 2 \pi / 3} \}_{n=0}^2$ and two
zero eigenvalues, with $a_0(T) = g_0(T) = 2$ giving index $\nu_0(T) = 1$.

\begin{figure}
\includegraphics[width=0.5\textwidth]{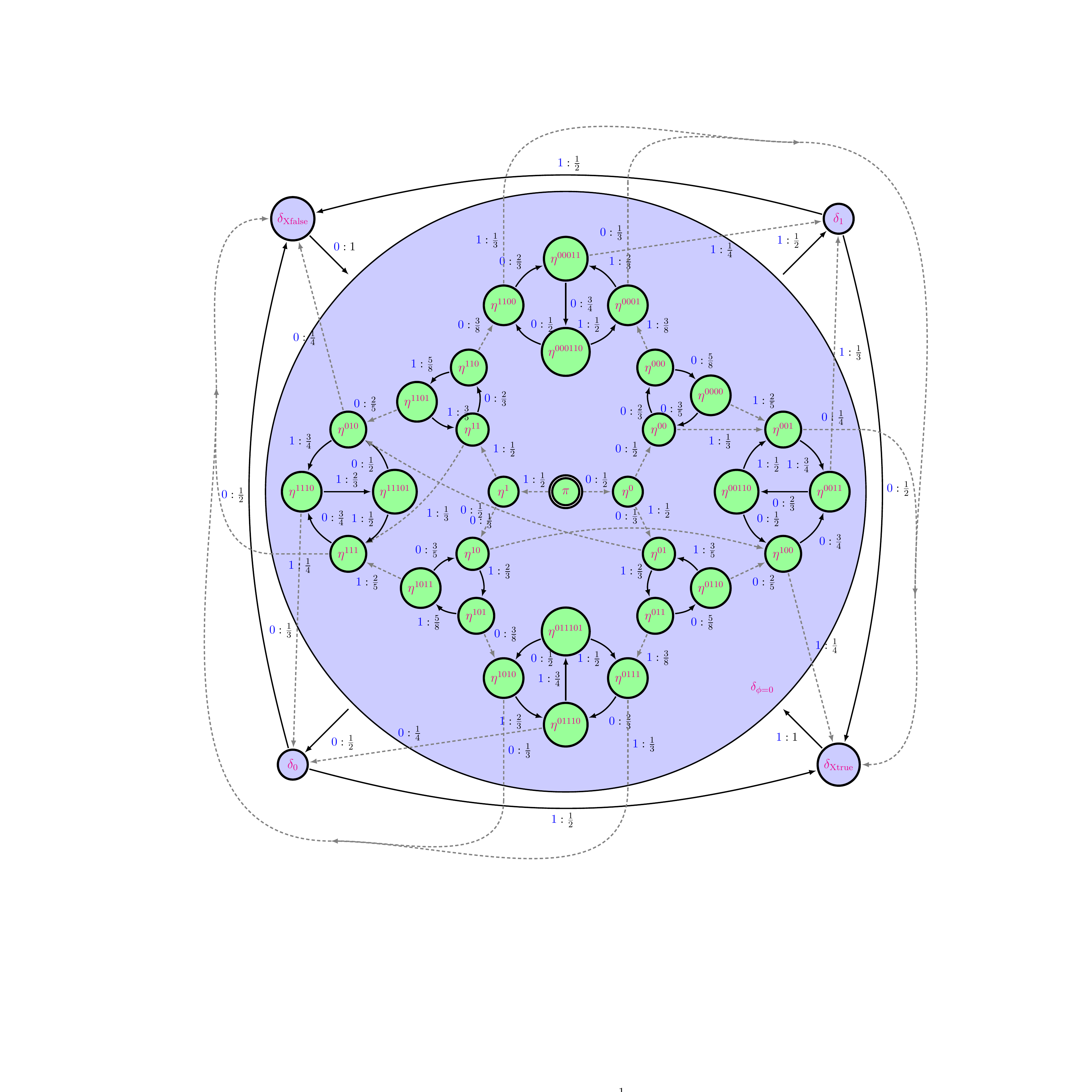}
\caption{MSP of the RRXOR Process' \eM: Grayed out (and dashed) transitions
	permanently leave the states from which they came. Recognizing the manner
	by which these transitions partition the mixed-state space allows
	simplified spectrum calculations. The directed graph structure is
	inherently nonplanar. The large blue recurrent state should be visualized
	as being \emph{behind} the transient states; it does not contain them.
	}
\label{fig:RRXMSP}
\end{figure}

The period-$3$ modulation leads to a phase ambiguity when an observer
synchronizes to the process, an ambiguity that resolved in the MSP transient
structure. This resolution is rather complicated, as made explicit in the RRXOR
Process' \syncMSP, shown in Fig.~\ref{fig:RRXMSP}. There are $31$ transient
states of uncertainty, in addition to the five recurrent states---$36$ causal
states in total.

Since we derived the \eM's \syncMSP, $W = \mathcal{W}$. Hence, the MSP's layout
depicts the information processing involved while an observer synchronizes to
the RRXOR Process. This graphically demonstrates the burden on an optimal
predictor, even one that only needs to learn an average of $\hmu$ bits per
observation to optimally predict the process.

The MSP introduces new, relevant zero eigenvalues associated with its transient
states. In particular, the first-encountered tree-like transients (starting
with mixed-state $\pi$) introduce new Jordan blocks up to dimension $2$.
Overall, the $0$-eigenspace of $W$ has index $2$, so that $\nu_0(W) = 2$.

Two different sets of leaky-period-$3$ structures appear in the MSP transients.
There are four leaky three-state cycles, each with the same leaky-period-3
contributions to the spectrum: $\bigl\{ (\tfrac{1}{4})^{1/3} e^{i n 2 \pi / 3}
\bigr\}_{n=0}^2$. There are also four leaky four-state cycles, each with a
leaky-period-3 contribution \emph{and} symmetry-breaking 0-eigenvalue
contribution to the spectrum: $\bigl\{ (\tfrac{1}{2})^{1/3} e^{i n 2 \pi / 3}
\bigr\}_{n=0}^2 \cup \{ 0\}$. The difference in eigenvalue magnitude,
$(\tfrac{1}{4})^{1/3}$ versus $(\tfrac{1}{2})^{1/3}$, implies different
timescales of synchronization associated with distinct learning tasks. For
example, an immediate lesson is that it takes longer (on average) to escape the
$4$-state leaky-period-$3$ components (from the time of arrival) than to escape
the preceding $3$-state leaky-period-$3$ components of the synchronizing
metadynamic.

\begin{figure}
\begin{center}
\begin{overpic}[width=0.5\textwidth,unit=1mm] 
      {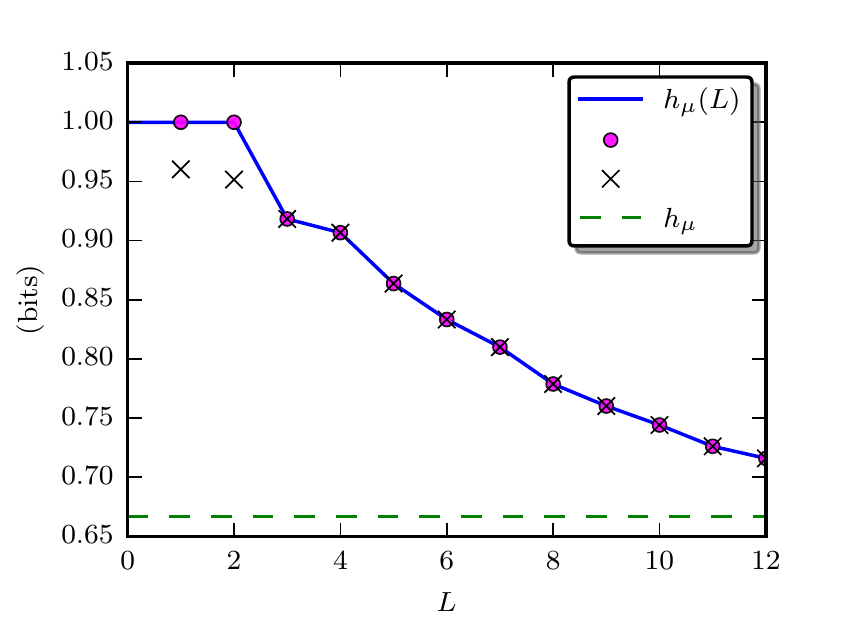}
  \put(67,50.5){\scalebox{0.85}{$\hEphemeral  \! +  \hPersistent$}}
  \put(70,46.5){\scalebox{0.85}{$\hPersistent$}}
\end{overpic}
\end{center}
\caption{Ephemeral ($\hEphemeral(L)$) and persistent ($\hPersistent(L)$)
	contributions to the myopic entropy rate ($\hmu(L)$). The ephemeral
	contribution lasts only up to $L = \nu_0(W) = 2$.
	}
\label{fig:RRXhmuL}
\end{figure}

\begin{figure}
\includegraphics[width=0.5\textwidth]{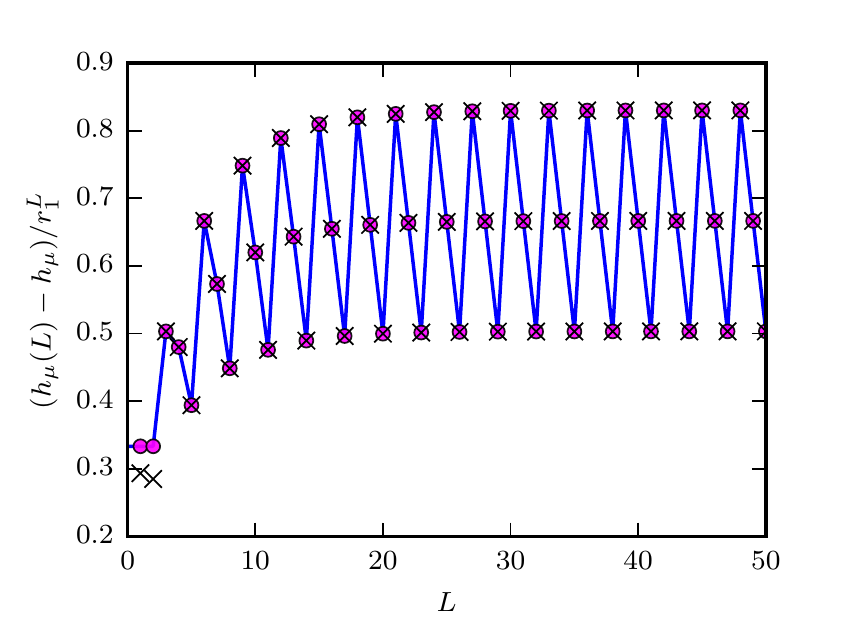}
\caption{Tails of the myopic entropy convergence $\hmu(L)$ shown in Fig.
	~\ref{fig:RRXhmuL} decay according to two different leaky period-three
	envelopes. The latter correspond to the two qualitatively different types
	of transient synchronization cycles in the MSP of Fig.~\ref{fig:RRXMSP}.
	One of the transient cycles has a relatively fast decay rate of $r_2 =
	(1/4)^{1/3}$.  While the slower decay rate of $r_1 = (1/2)^{1/3}$ dominates
	$\hmu(L)$'s deviation from $\hmu$ at large $L$.
	}
\label{fig:RRXhmuL_logscaled}
\end{figure}

The entropy rate convergence plots of Figs.~\ref{fig:RRXhmuL}
and~\ref{fig:RRXhmuL_logscaled} reveal a sophisticated predictability
modulation that simply could not have been gleaned from the spectra of
Fig.~\ref{fig:RRXPowerSpectrum}. Figure~\ref{fig:RRXhmuL_logscaled} emphasizes
the dominance of the slowest-decaying eigenmodes for large $L$. Such
oscillations under the exponential convergence to synchronization are typical.
However, as seen in comparison with Fig.~\ref{fig:RRXhmuL} much of the
uncertainty may be reduced before this asymptotic mode comes to dominate.
Ultimately, synchronization to optimal prediction may involve important
contributions from all modes of the mixed-state-to-state metadynamic.

This detailed analysis of the RRXOR Process suggests several general lessons
about how we view information in stochastic processes. First, as information
processing increases in sophistication, a vanishing amount of a process'
intrinsic structure will be discernible at low-orders of correlation. Second,
logical computation, as implemented by universal logic gates, primarily
operates above pairwise correlation. And so, finally, there is substantial
motivation to move beyond measures of pairwise correlation. We must learn to
recognize hidden structures and to use higher-order structural investigations
to better understand information processing. This is critical to empirically
probing functionality in biological and engineered processes.

\section{Conclusion}
\label{sec:Conclusion}

Surprisingly, many questions we ask about structured stochastic, nonlinear
processes implicate a linear dynamic over an appropriate hidden state space.
That is, there is an implied hidden Markov model. The promise is that once the
dynamic is found for the question of interest, one can make progress in
analyzing it. Unfortunately, a roadblock immediately arises: these hidden
linear dynamics are generically nondiagonalizable for questions related to
prediction and to information and complexity measures. Deploying Part I's
meromorphic functional calculus, though, circumvents the roadblock. Using it,
we determined closed-form expressions for a very wide range of information and
complexity measures. Often, these expressions turned out to be direct functions
of the HMM's transition dynamic.

This allowed us to catalog in detail the range of possible convergence
behaviors for correlation and myopic uncertainty. We then considered complexity
measures that accumulate during the transient relaxation to observer
synchronization. We also introduced the new notion of complexity spectra, gave
a new kind of information-theoretic signal analysis in terms of coronal
spectrograms, and highlighted common simplifications for special cases, such as
almost diagonalizable dynamics. We closed by analyzing several families of
finite and infinite Markov and cryptic order processes and emphasized the
importance of higher-than-pairwise-order correlations, showing how the excess
entropy spectrum is the key diagnostic tool for them.  

The analytical completeness might suggest that we have reached an end. Partly,
but the truth we seek is rather farther down the road. The meromorphic
functional calculus of nondiagonalizable operators merely sets the stage for
the next challenges---to develop complexity measures and structural
decompositions for infinite-state and infinite excess entropy processes.
Hopefully, the new toolset will help us scale the hierarchies of truly complex
processes outlined in Refs.  \cite{Crut01a,Crut15a,Marz17a}, at a minimum
giving exact answers at each stage of a convergent series of
finite-\eM\ approximations.

\section*{Acknowledgments}

The authors thank Alec Boyd, Chris Ellison, Ryan James, John Mahoney, and
Dowman Varn for helpful discussions. JPC thanks the Santa Fe Institute for its
hospitality. This material is based upon work supported by, or in part by, the
U. S. Army Research Laboratory and the U. S. Army Research Office under
contract numbers W911NF-12-1-0234, W911NF-13-1-0340, and W911NF-13-1-0390.

\appendix

\section{Example Analytical Calculations}
\label{sec:ExactCalculation}

To exercise the operational nature of the framework introduced, the following
explicitly carries out the analytic calculations to obtain the closed-form
complexity measures for the $(p, q)$-parametrized $(2\text{-}1)$-GP-$(2)$
Process. This process was already visually explored in the second column of
Table~\ref{table:infiniteR}. And so, the goal here is primarily
pedagogical---providing insight and better explicating particular calculational
steps. The appendix demonstrates a variety of techniques in the spirit of a
tutorial, though many were not called out in the main development.

\subsection{Process and spectra features}

\begin{figure}
\begin{center}
\includegraphics[width=0.3\textwidth]{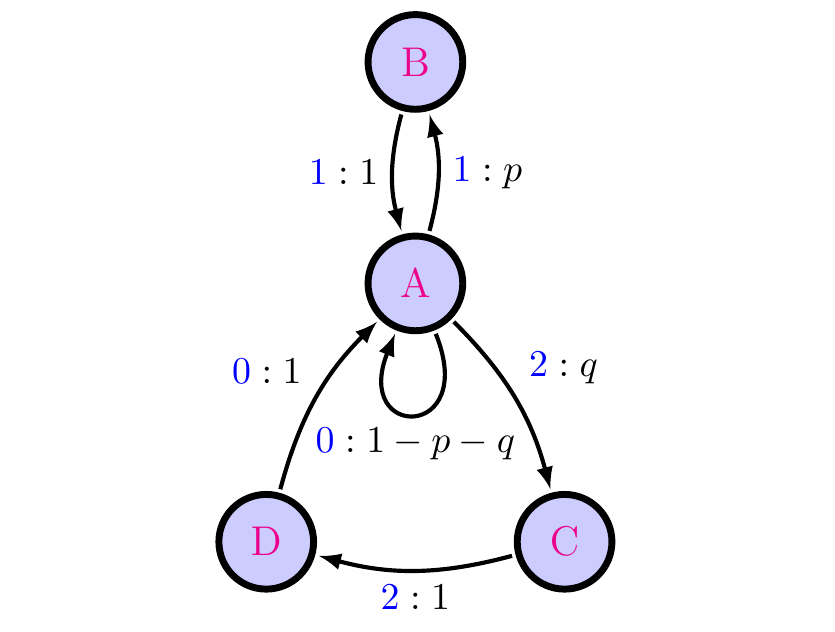}
\end{center}
\caption{\EM\ of the $(2\text{-}1)$-GP-$(2)$ Process.}
\label{fig:21GP2_eM}
\end{figure}

The $(p, q)$-parametrized $(2\text{-}1)$-GP-$(2)$ Process is described by its
\eM, whose state-transition diagram was shown in the first row and second
column of Table~\ref{table:infiniteR} and is reproduced here in Fig.
\ref{fig:21GP2_eM}. Formally, the $(2\text{-}1)$-GP-$(2)$ stationary stochastic
process is generated by the HMM $\mathcal{M}_{\epsilon \text{M}} = \bigl(
\SSet, \Abet, \{ T^{(\ms)} \}_{\ms \in \Abet}, \mxst_0 = \pi \bigr)$. That is,
$\mathcal{M}$ consists of a set of hidden causal states $\SSet = \{
{\color{magenta}\text{A}}, {\color{magenta}\text{B}},
{\color{magenta}\text{C}}, {\color{magenta}\text{D}} \}$, an alphabet $\Abet =
\{ {\color{blue}0}, {\color{blue}1}, {\color{blue}2} \}$ of symbols emitted to
form the observed process, and a set $\bigl\{ T^{(\ms)} : T^{(\ms)}_{s, s'} =
\Pr(X_t = \ms, \St_{t+1} = s' | \St_t = s ) \bigr\}_{\ms \in \Abet}$ of
symbol-labeled transition matrices. These are:
\begin{align*}
T^{(0)} &= 
\begin{bmatrix}
  \scalebox{0.75}{$1 \! - \! p \! - \! q$} & 0 & 0 & 0 \\
  0 & 0 & 0 & 0 \\
  0 & 0 & 0 & 0 \\
  1 & 0 & 0 & 0 \\
\end{bmatrix}
~, \\
T^{(1)} &= 
\begin{bmatrix}
  \; 0 \; &  p \, &  0 \, &  0 \; \\
  \; 1 & 0 & 0 & 0 \\
  \; 0 & 0 & 0 & 0 \\
  \; 0 & 0 & 0 & 0 \\
\end{bmatrix}
~, \\
\intertext{and}
T^{(2)} &= 
\begin{bmatrix}
  \; 0 \; &  0 \, &  q \, &  0 \; \\
  \; 0 & 0 & 0 & 0 \\
  \; 0 & 0 & 0 & 1 \\
  \; 0 & 0 & 0 & 0 \\
\end{bmatrix}
~.
\end{align*}
The symbol-labeled transition matrices sum to the row-stochastic internal
state-to-state transition matrix: 
\begin{align*}
T = \sum_{\ms \in \Abet} T^{(\ms)} =
\begin{bmatrix}
  \scalebox{0.75}{$1 \! - \! p \! - \! q$}  &  p &  q  &  0  \\
   1 & 0 & 0 & 0 \\
   0 & 0 & 0 & 1 \\
   1 & 0 & 0 & 0 \\
\end{bmatrix}
~.
\end{align*}
This is a Markov chain over the hidden states. From it we find the stationary
state distribution $\bra{\pi} = \bra{\pi} T$:
\begin{align*}
\bra{\pi} = \frac{1}{1+p+2q} 
\begin{bmatrix}
  1 & p & q & q 
\end{bmatrix} ~.
\end{align*}

The first analysis task is to determine the eigenvalues and associated
projection operators for the internal state-to-state transition matrix $T$.
From:
\begin{align*}
\det (T-\lambda I)
  & = \lambda (\lambda - 1) \bigl( \lambda^2 + (p+q) \lambda + q \bigr) \\
  & = 0
  ~, 
\end{align*}
we find the four eigenvalues:
\begin{align*}
\Lambda_T = \bigl\{ 1, \, 0,
  \, - \tfrac{1}{2}(p+q) \pm \tfrac{1}{2} \sqrt{(p+q)^2 - 4q} \, \bigr\}
  ~.
\end{align*}

\begin{figure}
\begin{center}
\includegraphics[width=0.42\textwidth]{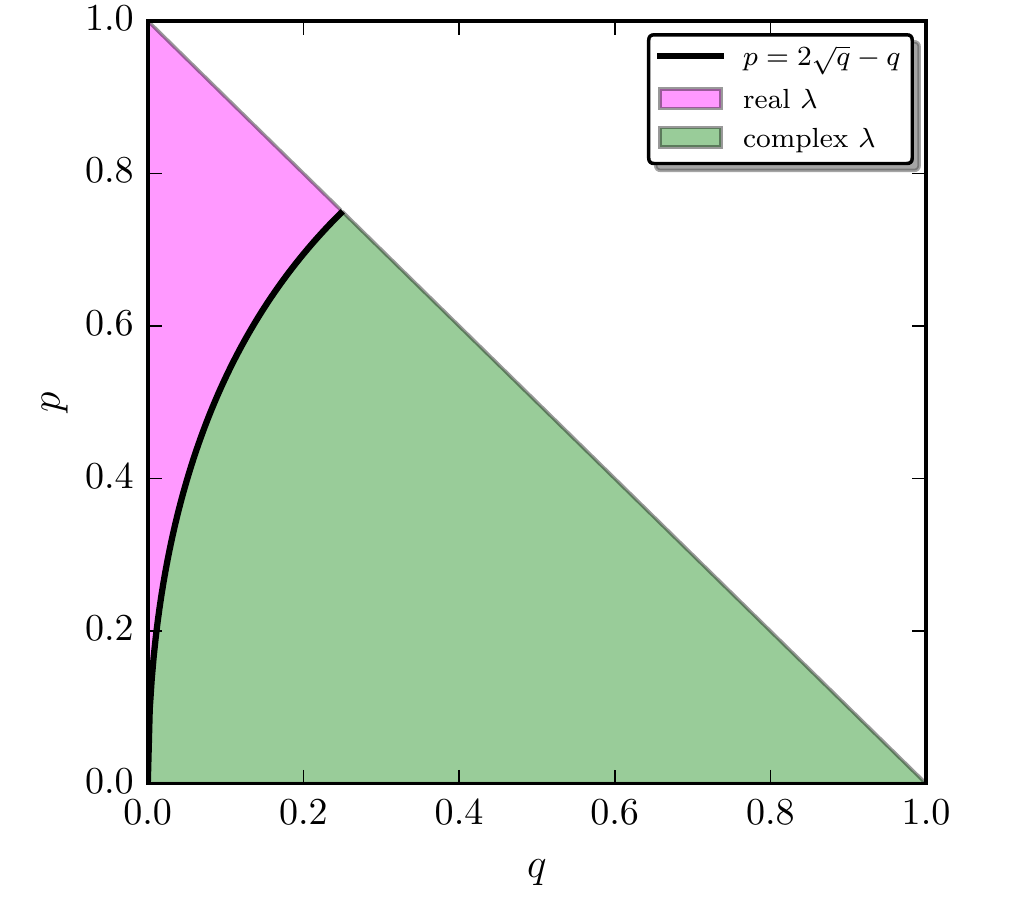}
\end{center}
\caption{Transition-probability space for the $(p, q)$-parametrized
  $(2\text{-}1)$-GP-$(2)$ Process.
  }
\label{fig:TranProbSpace}
\end{figure}

All eigenvalues are real for $p \geq 2 \sqrt{q} - q$. Two are complex with nonzero imaginary part when $p < 2 \sqrt{q} - q$. Putting this together with the transition-probability consistency constraint that $p+q < 1$ yields the map of the transition-parameter space shown in Fig. \ref{fig:TranProbSpace}.

For a generic choice of the parameter setting $(p, q)$, all $T$'s eigenvalues
are unique, and so $T$ is diagonalizable. However, two of the eigenvalues
become degenerate along the parameter subspace $p = 2 \sqrt{q} - q$, giving
$\Lambda_T = \{ 1, 0, -\sqrt{q} \}$. We find that the algebraic multiplicity
$a_{-\sqrt{q}} = 2$ is larger than the geometric multiplicity $g_{-\sqrt{q}} =
1$, yielding nondiagonalizability ($\nu_{-\sqrt{q}} = 2$) along this ($p = 2
\sqrt{q} - q$)-submanifold. More broadly, experience has shown that eigenvalues
generically induce nondiagonalizability when they collide and scatter in the
complex plane. For example, this occurs when a pair of eigenvalues first
``entangle'' to become complex conjugate pairs.

For any parameter setting ($p, q$), we find $T$'s right eigenvectors from $(T -
\lambda I) \ket{\lambda} = \ket{0}$. They are:
\begin{align*}
\ket{0} & = 
	\begin{bmatrix}
		0 & 1 & -p/q & 0
	\end{bmatrix}^\top \\
\intertext{and}
\ket{\lambda} & = 
	\begin{bmatrix}
		\lambda^2 & \lambda & 1 & \lambda
	\end{bmatrix}^\top
  ~,
\end{align*}
for all $\lambda \in \Lambda_T \setminus \{ 0 \}$. Similarly, we find the left eigenvectors of $T$ from $\bra{\lambda} (T - \lambda I) = \ket{0}$:
\begin{align*}
\bra{0} &= 
	\begin{bmatrix}
	0 & 1 & 0 & -1
	\end{bmatrix} \\
\intertext{and}
\bra{\lambda} &= 
	\begin{bmatrix}
	\lambda^2 & p \lambda & q \lambda  & q
	\end{bmatrix}
  ~,
\end{align*}
for all $\lambda \in \Lambda_T \setminus \{ 0 \}$. Clearly, the left
eigenvectors are \emph{not} simply the complex-conjugate transpose of the right
eigenvectors. This is a signature of the more intricate algebraic structure in
these processes.

Since $T$ is generically diagonalizable, all of the projection operators are
simply the normalized ket-bra outer products:
\begin{align*}
T_\lambda = \frac{ \ket{\lambda} \bra{\lambda} }{ \braket{\lambda | \lambda} } ~,
\end{align*}
so long as $p \neq 2 \sqrt{q} - q$. To wit:
\begin{align*}
T_1 & = \frac{\ket{1} \bra{1}}{\braket{1 | 1}} \\
    & = \ket{\one} \bra{\pi}
  ~.
\end{align*}

Along the nondiagonalizable ($p = 2 \sqrt{q} - q$)-subspace of parameter
settings: $\braket{^-\sqrt{q} | ^- \sqrt{q}} = 0$. This corresponds to the fact
that the right and left eigenvectors are now dual to the left and right
\emph{generalized} eigenvectors, rather than being dual to each other. We find
the projection operator for the nondiagonalizable eigenspace is:
\begin{align*}
T_{-\sqrt{q}} = \frac{ \ket{ ^-\sqrt{q}^{(1)} } \bra{ ^-\sqrt{q}^{(2)} } }{ \braket{ ^-\sqrt{q}^{(2)} | ^-\sqrt{q}^{(1)} } } + 
\frac{ \ket{ ^-\sqrt{q}^{(2)} } \bra{ ^-\sqrt{q}^{(1)} } }{ \braket{ ^-\sqrt{q}^{(1)} | ^-\sqrt{q}^{(2)} } }
  ~,
\end{align*}
where:
\begin{align*}
\ket{ ^-\sqrt{q}^{(1)} }
  & = \begin{bmatrix}
	q & -\sqrt{q} & \; 1 & -\sqrt{q}
	\end{bmatrix}^\top ~, \\
\ket{ ^-\sqrt{q}^{(2)} }
  & = \begin{bmatrix}
	-\sqrt{q} & 0 & 1/\sqrt{q} & 0
	\end{bmatrix}^\top ~, \\
\bra{ ^-\sqrt{q}^{(1)} }
  & = \begin{bmatrix}
	q & q (\sqrt{q} - 2) & -q \sqrt{q} & q
	\end{bmatrix} ~, \text{ and} \\
\bra{ ^-\sqrt{q}^{(2)} }
  & = \begin{bmatrix}
	-\sqrt{q} & 0 & 0 & \sqrt{q}
	\end{bmatrix}
	~.
\end{align*}
Equivalently, $T_{-\sqrt{q}} = I - T_1 - T_0$. The projection operators for the
remaining ($\lambda = 0, 1$) eigenspaces retain the same form as before:
$T_\lambda = \ket{\lambda} \bra{\lambda} / \braket{\lambda | \lambda}$.

\subsection{Observed correlation}

The pieces are now in place to calculate the observable correlation and power spectrum. Recall that we derived the general spectral decomposition of the 
autocorrelation function $\gamma(L) = \left\langle \overline{\MS}_t \MS_{t+L}
\right\rangle_t$:
\begin{align*} 
\gamma(L) & = \sum_{\lambda \in \Lambda_T \atop \lambda \neq 0} 
  \sum_{m = 0}^{\nu_\lambda -1} \corrbra  T_{\lambda, m} \corrket
   \binom{| L |-1}{m}  \lambda^{| L |-1-m} \nonumber \\
  & \qquad + [0 \in \Lambda_{T}] 
   \sum_{m = 0}^{\nu_0 -1} \corrbra  T_{0} T^{m}  \corrket
   \delta_{| L |-1, m}
\end{align*} 
for nonzero integer $L$,
where: 
\begin{align*}
\corrbra & = \bra{\pi} \sum_{\ms \in \Abet} \overline{\ms} T^{(\ms)} \\
  & = \frac{1}{1+p+2q} 
	\begin{bmatrix}
	p & p & 2q & 2q
	\end{bmatrix}
  ~,
\end{align*}
and: 
\begin{align*}
\corrket & = \sum_{\ms \in \Abet}  \ms T^{(\ms)}  \ket{\one} \\
  & = \begin{bmatrix}
	p+2q & 1 & 2 & 0
	\end{bmatrix}^\top
  ~.
\end{align*}
For generic parameter settings, this reduces to:
\begin{align*} 
\gamma(L) & = 
  \frac{\corrbra  0 \rangle \langle 0  \corrket }{\braket{0 | 0} }
   \delta_{| L |, 1} \\
   & \qquad + \sum_{\lambda \in \Lambda_T \atop \lambda \neq 0} 
   \frac{ \corrbra  \lambda \rangle \langle \lambda \corrket }
   {\braket{\lambda | \lambda} } \lambda^{| L |-1} \\
  & = 4 \left( \frac{p + 2q}{1+p+2q} \right)^2 + \frac{ -p }{ 1+p+2q }
   \delta_{| L |, 1} \\
   & \qquad + 
  \frac{p + 2q}{1+p+2q} \sum_{\lambda \in \Lambda_T \atop \lambda \neq 0, 1}
  \frac{(\lambda+1)^2 (p \lambda + 2 q) }{\lambda^3 + p \lambda + 2 q} 
  \lambda^{| L |-1}
  ~.
\end{align*} 
Notably, the autocorrelation splits into an ephemeral part (via the Kronecker
delta) due to the zero eigenvalue, an exponentially decaying oscillatory part
due to the two eigenvalues with magnitude between $0$ and $1$, and an
asymptotic part that survives even as $L \to \infty$ due to the eigenvalue of
unity. Moreover, for $p < 2 \sqrt{q} - q$, the two nontrivial eigenvalues
become complex conjugate pairs. This allows us to rewrite the autocorrelation
for the $(2\text{-})$-GP-$(2)$ process for $|L| \geq 2$ concisely as:
\begin{align*}
\gamma(L) 
&= 
\beta^2 +
\beta \, q^{|L| / 2} \, \text{Re} \bigl( \zeta \, e^{i \omega_\xi |L| } \bigr)
 ~,
\end{align*}
where:
\begin{align*}
\beta & \equiv \frac{2(p + 2q)}{1+p+2q} ~,  \\
\zeta & \equiv \frac{(\xi +1)^2 (p \xi + 2 q) }{\xi ( \xi^3 + p \xi + 2 q ) }
  ~, \\
\xi & \equiv -\frac{1}{2} (p+q) + i \frac{1}{2} \sqrt{4q - (p+q)^2}
  ~, \text{ and} \\
\omega_\xi & \equiv \frac{\pi}{2}
  + \arctan \bigl( \frac{p+q}{\sqrt{4q - (p+q)^2}} \bigr)
  ~.
\end{align*}
The latter form reveals that the magnitude of the largest nonunity eigenvalue
$|\xi| = \sqrt{q}$ controls the slowest rate of decay of observed correlation
$|\gamma(L)| \sim \sqrt{q}^L$ and that the complex phase of the eigenvalues
determine the oscillations within this exponentially decaying envelope.

We found the general spectral decomposition of the continuous part of the power
spectrum is:
\begin{align*}
P_\text{c}(\omega) & =  \bigl\langle \left| x \right|^2 \bigr\rangle + 2 
    \sum_{\lambda \in \Lambda_T} \sum_{m = 0}^{\nu_\lambda - 1} \, \text{Re}  \,
  \frac{ \corrbra T_{\lambda,m} \corrket  }{(e^{i \omega} - \lambda)^{m+1}} 
   ~.
\end{align*}

Also, recall from earlier that the delta functions of the power spectrum arise
from $T$'s eigenvalues that lie on the unit circle:
\begin{align*}
P_\text{d}(\omega) & =  
    \sum_{k = -\infty}^{\infty} 
  \sum_{\lambda \in \Lambda_T \atop |\lambda| = 1}
  2 \pi \, \delta( \omega - \omega_\lambda + 2 \pi k) 
  \nonumber \\
  & \qquad \qquad \qquad \times 
  \text{Re} \bigl( \lambda^{-1} \corrbra \, T_\lambda \corrket \bigr)
 ~,
\end{align*}
where $\omega_\lambda$ is related to $\lambda$ by $\lambda = e^{i
\omega_\lambda}$.
As long as $p+q < 1$, $\lambda=1$ is the only eigenvalue that lies on the unit circle,
so that:
\begin{align*}
P_\text{d}(\omega)  =  \sum_{k = -\infty}^{\infty} 
    2 \pi \, \delta( \omega + 2 \pi k) \text{Re}  \corrbra \, T_1 \corrket
	~,
\end{align*}
with:
\begin{align*}
\text{Re}  \corrbra \, T_1 \corrket & = \beta^2 \\
  & = 4 \left( \frac{p + 2q}{1+p+2q} \right)^2
  ~.
\end{align*}

Putting this all together we find the complete power spectrum for the
$(2\text{-}1)$-GP-$(2)$ Process:
\begin{align*}
P(\omega) & =
  \frac{8q}{1+p+2q} +  \frac{2p}{1+p+2q} \bigl[ 1 - \cos(\omega) \bigr] \\
  & \qquad + \beta \sum_{\lambda \in \Lambda_T \atop \lambda \neq 0, 1}
  \text{Re} \left( \frac{(\lambda+1)^2 (p \lambda + 2 q) }
  {\lambda^3 + p \lambda + 2 q} \cdot \frac{1}{e^{i \omega} - \lambda }
   \right) \\
   & \qquad + 2 \pi \beta^2 \sum_{k = -\infty}^{\infty} \delta( \omega + 2 \pi k)  ~,
\end{align*}
Thus, the process exhibits delta functions from the eigenvalue on the unit
circle, continuous Lorentzian-like line profiles emanating from finite
eigenvalues inside the unit circle which express the process' chaotic nature,
and the unique sinusoidal contribution that can only come from zero
eigenvalues.  

Whenever $p < 2 \sqrt{q} - q$, the two nontrivial eigenvalues become complex
conjugate pairs. This allows us to rewrite the process' power spectrum in a
more transparent way:
\begin{align*}
P(\omega) 
  & =  \frac{8q}{1+p+2q} + \frac{2p}{1+p+2q}
    \bigl[ 1 - \cos(\omega) \bigr] \\
  & \qquad + \beta \, \text{Re}
  \left( \frac{\zeta \xi}{e^{i \omega} - \xi } + \frac{\zeta \xi}{e^{- i
  \omega} - \xi } \right) \\
  & \qquad + 2 \pi \beta^2 \sum_{k = -\infty}^{\infty} \delta( \omega + 2 \pi k)
  ~,
\end{align*}
which is clearly symmetric about $\omega = 0$.

The level of completeness achieved is notable: we calculated these properties
exactly in closed form for this infinite-Markov order stochastic process over
the full range of possible parameter settings. Moreover, once it constructs the process' \syncMSP, the next section goes on to produce the same level of analytical completeness but for predictability.

\subsection{Predictability}

To analyze the process' predictability, we need the \syncMSP\ of any of its
generators. Since we started already with the \eM, we directly determine its
\syncMSP. This gives the mixed-state transition matrix $W = \mathcal{W}$ that,
in turn, suffices for calculating both predict\emph{ability} in this section
and the synchronization necessary for predic\emph{tion} in the next.

\begin{figure}
\begin{center}
\includegraphics[width=0.42\textwidth]{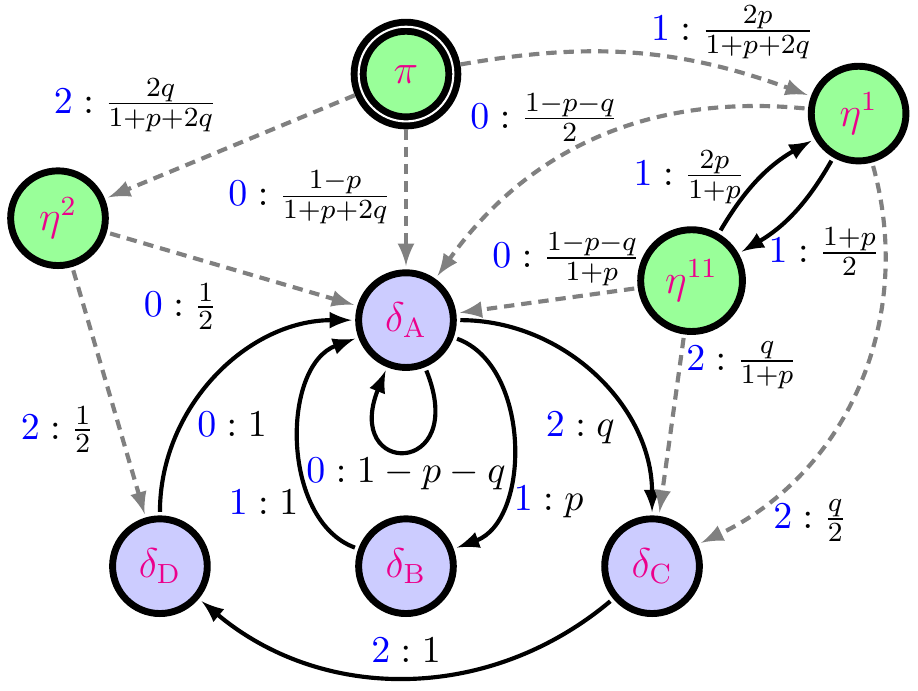}
\end{center}
\caption{\syncMSP\ of the $(2\text{-}1)$-GP-$(2)$ Process.}
\label{fig:syncMSP}
\end{figure}

We construct the \syncMSP\ by calculating all mixed states that can be induced
by observation from the start distribution $\pi$ and then calculating the
transition probabilities between them. There are eight such mixed-state
distributions over $\SSet$. However, four of them (${\color{magenta}
\delta_\text{A}}$, ${\color{magenta} \delta_\text{B}}$, ${\color{magenta}
\delta_\text{C}}$, and ${\color{magenta} \delta_\text{D}}$) correspond to
completely synchronized peaked distributions. The other four states are new
(relative to the recurrent states) transient states and correspond to transient
states of recurrent-state uncertainty during synchronization. Calculating, we
find the eight unique mixed-state distributions iteratively from:
\begin{align*}
\bra{\mxst^{wx}} = \frac{ \bra{\mxst^{w}} T^{(x)}  }{ \bra{\mxst^{w}} T^{(x)} \ket{\one} }
  ~,
\end{align*}
starting with:
\begin{align*}
\bra{\mxst^{x}} = \frac{ \bra{ \pi } T^{(x)}  }{ \bra{ \pi } T^{(x)} \ket{\one} }
  ~,
\end{align*}
are:
\begin{align*}
\bra{\pi} & = \frac{1}{1+p+2q} 
	\begin{bmatrix}
  	1 & p & q & q 
	\end{bmatrix} ~, \\
\bra{\mxst^{1}} & = \frac{1}{2} 
	\begin{bmatrix}
  	1 & 1 & 0 & 0 
	\end{bmatrix} ~, \\
\bra{\mxst^{11}} & = \frac{1}{1+p} 
	\begin{bmatrix}
  	1 & p & 0 & 0 
	\end{bmatrix} ~, \\
\bra{\mxst^{2}} & = \frac{1}{2} 
	\begin{bmatrix}
  	0 & 0 & 1 & 1 
	\end{bmatrix} ~, \\
\bra{ \delta_\text{A}} & = 
	\begin{bmatrix}
  	1 & 0 & 0 & 0 
	\end{bmatrix} ~, \\
\bra{ \delta_\text{B}} & = 
	\begin{bmatrix}
  	0 & 1 & 0 & 0 
	\end{bmatrix} ~, \\
\bra{ \delta_\text{C}} & = 
	\begin{bmatrix}
	0 & 0 & 1 & 0 
	\end{bmatrix} ~, \text{ and} \\
\bra{ \delta_\text{D}} & = 
	\begin{bmatrix}
  	0 & 0 & 0 & 1 
	\end{bmatrix}
  ~.
\end{align*}
In this, the transient mixed states $\mxst^{w}$ are labeled according to the
shortest word $w$ that induces them. These distributions constitute the set of
mixed-states $\MxSSet$ of the \syncMSP. Moreover, each transition probability
from mixed state $\bra{\mxst^w}$ to mixed state $\bra{\mxst^{wx}}$ is
calculated as: $\bra{\mxst^w} T^{(x)} \ket{\one}$. Altogether, these
calculations yield the \syncMSP\ of the $(2\text{-}1)$-GP-$(2)$ Process,
reproduced in Fig. \ref{fig:syncMSP} from Table~{\ref{table:infiniteR}} for
convenience.

As a HMM, the \eM's \syncMSP\ is specified by the $4$-tuple:
$\mathcal{M}_\text{\syncMSP} = \bigl( \MxSSet, \, \Abet, \, \{
\mathcal{W}^{(\ms)} \}_{\ms \in \Abet}, \, \mu_0 \! = \! \delta_\pi \bigr)$,
where $\MxSSet$ is the set of mixed states just quoted, $\Abet$ is the same
observable alphabet as before, $\bigl\{ \mathcal{W}^{(\ms)} \bigr\}_{\ms \in
\Abet}$ is the set of symbol-labeled transition matrices among the mixed
states, and $\StartMS = \begin{bmatrix} 1 & 0 & 0 & 0 & 0 & 0 & 0 & 0
\end{bmatrix}$ is the start distribution over the mixed states.

With this new linear metadynamic in hand, our next step is to calculate the eigenvalues and projection operators of the internal mixed-state-to-state transition dynamic $\mathcal{W} = \sum_{\ms \in \Abet} \mathcal{W}^{(\ms)}$.
$\mathcal{W}$ can be explicitly represented in the block-matrix form:
\begin{align*}
\mathcal{W} = \begin{bmatrix} A & B \\ \boldsymbol{0} & T \end{bmatrix}
  ~,
\end{align*}
where:
\begin{align*}
A & = \begin{bmatrix} 
0 & \tfrac{2p}{1+p+2q} & 0 & \tfrac{2q}{1+p+2q} \\
0 & 0 & \tfrac{1+p}{2} & 0 \\
0 & \tfrac{2p}{1+p} & 0 & 0 \\
0 & 0 & 0 & 0 
\end{bmatrix}
  ~ \text{and} \\
B & = \begin{bmatrix} 
\tfrac{1-p}{1+p+2q} & 0 & 0 & 0 \\
\tfrac{1-p-q}{2} & 0 & \tfrac{q}{2} & 0 \\
\tfrac{1-p-q}{1+p} & 0 & \tfrac{q}{1+p} & 0 \\
\tfrac{1}{2} & 0 & 0 & \tfrac{1}{2} 
\end{bmatrix}
  ~,
\end{align*}
and $T$ is the same as the state-to-state internal transition matrix of the \eM\ from earlier.

$\mathcal{W}$'s eigenvalues are thus relatively straightforward to calculate
since $\text{det}(\mathcal{W} - \lambda I) = \text{det}(A - \lambda I)
\text{det}(T - \lambda I)$ implies that:
\begin{align*}
\Lambda_\mathcal{W} = \Lambda_A \cup \Lambda_T ~.
\end{align*}
The new eigenvalues introduced by the feedback matrix $A$ are $\Lambda_A = \{
0, \pm \sqrt{p} \}$ with $\nu_0=2$. While the other eigenvalues ($\pm
\sqrt{p}$), found most easily from Part I's cyclic eigenvalue rule,
are associated with diagonalizable subspaces.
It is important to note that, while $T$ was only nondiagonalizable along a very
special submanifold in parameter space, the mixed-state-to-state metadynamic is
\emph{generically} nondiagonalizable over all parameter settings. This
nondiagonalizability corresponds to a special kind of symmetry breaking of
uncertainty during synchronization.

$\mathcal{W}$'s eigenvectors are most easily found through a two-step process.
Specifically, $\ket{^\pm \sqrt{p}_A}$ and $\bra{^\pm \sqrt{p}_A}$ (the
solutions of $A \ket{^\pm \sqrt{p}_A} = \pm \sqrt{p} \ket{^\pm \sqrt{p}_A}$ and
$\bra{^\pm \sqrt{p}_A} A = \pm \sqrt{p} \bra{^\pm \sqrt{p}_A}$) are found
first, and the result is used to reduce the number of unknowns when solving the
full eigenequations ($\mathcal{W} \ket{^\pm \sqrt{p}_\mathcal{W}} = \pm
\sqrt{p} \ket{^\pm \sqrt{p}_\mathcal{W}} $) for $ \ket{^\pm
\sqrt{p}_\mathcal{W}}$ and $\bra{^\pm \sqrt{p}_\mathcal{W}}$. Similarly, we can
recycle the restricted eigenvectors $\ket{\lambda_T}$ and $\bra{\lambda_T}$
found earlier for the \eM\ to reduce the number of unknowns when solving the
more general eigenvector problems for $\ket{\lambda_\mathcal{W}}$ and
$\bra{\lambda_\mathcal{W}}$ in cases where $\lambda \in \Lambda_T$. Performing
such a calculation, we find:
\begin{align*}
\ket{^\pm \sqrt{p}_\mathcal{W}}
  & = \begin{bmatrix}
\frac{1}{1+p+2q} & \tfrac{\pm \sqrt{p}}{2p} & \tfrac{1}{1+p} & 0 & 0 & 0 & 0 & 0
\end{bmatrix}^\top ~, \\
\bra{^\pm \sqrt{p}_\mathcal{W}}
  & = \begin{bmatrix}
	0 & ^\pm \sqrt{p} & \frac{1+p}{2} & 0 & -\tfrac{1 \pm \sqrt{p}}{2} & -\tfrac{p \pm \sqrt{p}}{2}  & 0 & 0
\end{bmatrix} ~, \\
\bra{\lambda_\mathcal{W}}
  & = \begin{bmatrix}
	0 & 0 & 0 & 0 &
	\lambda^2 & p \lambda & q \lambda  & q
	\end{bmatrix}
\text{ for all } \lambda \in \Lambda_T \setminus \{ 0 \} ~, \\
\ket{1_\mathcal{W}}
  & = 
	\begin{bmatrix}
	1 & 1 & 1 & 1 & 1 & 1 & 1 & 1
	\end{bmatrix}^\top ~, \text{ and} \\
\ket{\lambda_\mathcal{W}}
  & = \begin{bmatrix}
	0 &
	\frac{ \lambda (1+\lambda)}{2} &
	\frac{ \lambda (1+\lambda)}{1+p} &
	\frac{1+\lambda}{2} &
	\lambda^2 & \lambda & 1 & \lambda
	\end{bmatrix}^\top
	~,
\end{align*}
for $\lambda \in \Lambda_T \setminus \{ 0, 1 \}$. Moreover, $\mathcal{W}$'s
eigenvectors and generalized eigenvector corresponding to eigenvalue $0$ are: 
\begin{align*}
\ket{0_1^{(1)}} & = \StartMS^\top
  = \begin{bmatrix} 1 & 0 & 0 & 0 & 0 & 0 & 0 & 0
	\end{bmatrix}^\top ~, \\
\ket{0_1^{(2)}} & = \begin{bmatrix} 0 & 0 & 0 & \frac{1+p+2q}{2q} & 0 & 0 & 0 &
0 \end{bmatrix}^\top ~, \\
\ket{0_2^{(1)}} & =  \begin{bmatrix} 0 & \frac{1}{2} & \frac{p}{1+p} &
\frac{-p}{2q} & 0 & 1 & \frac{-p}{q} & 0 \end{bmatrix}^\top ~, \\
\bra{0_1^{(1)}} & =  \begin{bmatrix} 0 & 0 & 0 & 1 & 0 & 0 & \frac{-1}{2} &
\frac{-1}{2}  \end{bmatrix} ~, \\
\bra{0_1^{(2)}} & =  \begin{bmatrix} \frac{1+p+2q}{2q} & 0 & \frac{-(1+p)}{2q}
& 0 & 0 & 0 & \frac{-1}{2} & \frac{-1}{2} \end{bmatrix} ~, \text{ and} \\
\bra{0_2^{(1)}} & = \begin{bmatrix} 0 & 0 & 0 & 0 & 0 & 1 & 0 & -1 \end{bmatrix}
  ~.
\end{align*}
Above, we used the notation $\ket{0_k^{(m)}}$ for indexing generalized eigenvectors introduced in Part I.

All nondegenerate eigenvalues have projection operators of the form:
\begin{align*}
\mathcal{W}_\lambda = \frac{\ket{\lambda} \bra{\lambda}}{\braket{\lambda | \lambda}}
  ~.
\end{align*}
However, the degenerate and nondiagonalizable subspace associated with the zero
eigenvalue has the composite projection operator:
\begin{align*}
\mathcal{W}_0 &= 
\frac{\ket{0_1^{(1)}} \bra{0_1^{(2)}}}{\braket{0_1^{(2)} | 0_1^{(1)}}} + 
\frac{\ket{0_1^{(2)}} \bra{0_1^{(1)}}}{\braket{0_1^{(1)} | 0_1^{(2)}}} + 
\frac{\ket{0_2^{(1)}} \bra{0_2^{(1)}}}{\braket{0_2^{(1)} | 0_2^{(1)}}} ~.
\end{align*}

The fact that $\StartMS \lambda_\mathcal{W} \rangle = 0$ for all $\lambda \in \Lambda_T \setminus \bigl( \{ 1 \} \cup \Lambda_A \bigr)$ is an instantiation of a general result that greatly simplifies the calculations relating to predictability and prediction.

The remaining piece for analyzing predictability is the vector of
transition-entropies $\HWA$. A simple calculation, utilizing the fact that:
\begin{align*}
-\sum_i \frac{n_i}{d} \log \left( \frac{n_i}{d} \right)
   = \log(d) - \frac{1}{d} \sum_i n_i \log n_i 
  ~,
\end{align*}
when $\sum_i n_i = d$, yields:
\begin{widetext}
\begin{align*}
\HWA = \begin{bmatrix}
\log(1+p+2q) - \frac{1}{1+p+2q} \bigl[ 2q \log( 2q ) + 2p \log( 2p ) + (1-p) \log( 1-p ) \bigr] \\
1 - \frac{1}{2} \bigl[ q \log( q ) + (1+p) \log( 1+p ) + (1-p-q) \log( 1-p-q ) \bigr] \\
\log(1+p) - \frac{1}{1+p} \bigl[ q \log( q ) + 2p \log( 2p ) + (1-p-q) \log( 1-p-q ) \bigr] \\
1\\
- \bigl[ q \log( q ) + p \log( p ) + (1-p-q) \log( 1-p-q ) \bigr] \\
0 \\
0 \\
0 
\end{bmatrix} ~,
\end{align*}
where $\log$ is understood to be the base-2 logarithm $\log_2$.
\end{widetext}

Putting this all together, we can now calculate in full detail the myopic entropy rate $\hmu(L)$ that results from modeling the infinite-order $(2\text{-}1)$-GP-$(2)$ process as an order-$(L\text{-}1)$ Markov process:
\begin{align*}
\hmu(L) & = 
  \sum_{\lambda \in \Lambda_W \atop \lambda \neq 0}
  \sum_{m = 0}^{\nu_\lambda -1} \StartMS W_{\lambda, m} \HWA
  \binom{L-1}{m} \lambda^{L-1-m} \\ 
  & \quad +  \left[ 0 \in \Lambda_W \right] 
	\sum_{m=0}^{\nu_0 - 1} \delta_{L-1, m} 
	\StartMS W_0 W^m \HWA \\
  & = 
  \delta_{L,1} \StartMS  W_0  \HWA + \delta_{L, 2} \StartMS W W_0 \HWA \\ 
  & \quad + \StartMS  W_1  \HWA
  + \! \! \! \! \!  \sum_{\lambda \in \Lambda_W \setminus \Lambda_T}
  \! \! \! \! \!  \StartMS W_{\lambda} \HWA \lambda^{L-1} 
  .
\end{align*}
This oscillates under an exponentially decaying envelope 
as it approaches its asymptotic value of: 
\begin{align*}
\hmu & = \StartMS  W_1  \HWA \\
     & = \StartMS  1_\mathcal{W} \rangle \langle \pi_\mathcal{W}  \HWA \\
     & = \langle \pi_\mathcal{W}  \HWA = \braket{ \pi_T | \H(T^{\Abet}) } \\
     & = \frac{-q \log q - p \log p - (1-p-q) \log (1-p-q) }{1+p+2q}
	 ~.
\end{align*}
Simplifying the terms in the myopic entropy rate yields:
\begin{align*}
\StartMS W_0 \HWA
  & = \langle \delta_{\pi} \HWA \\
  & \qquad - \frac{1+p}{1+p+2q} \langle \delta_{\mxst^{11}} \HWA
\end{align*}
and:
\begin{align*}
\StartMS W W_0 \HWA = \frac{2q}{1+p+2q}
\end{align*}
for the two ephemeral contributions.

For $L \geq 3$, we find for odd $L$:
\begin{align*}
\hmu(L) - \hmu & = 
\frac{-p \log p + (1+p) \log(1+p) - 2p}{\sqrt{p} (1+p+2q)} \, p^{L/2}
\end{align*}
and for even $L$:
\begin{align*}
\hmu(L) - \hmu & = 
\frac{p \log p - (1+p) \log(1+p) + 2}{ 1+p+2q } \, p^{L/2}
  ~.
\end{align*}
This highlights the period-2 nature of the asymptotic decay.

The total mutual information between the observable past and observable future,
the excess entropy, is:
\begin{align*}
\EE & = \!\!\! \sum_{\lambda \in \Lambda_W \setminus \{ 1 \} } 
    \sum_{m=0}^{\nu_\lambda - 1} 
    \frac{1}{(1-\lambda)^{m+1}} 
    \StartMS  W_{\lambda, m}  \HWA \\
  & = \frac{(1-p-q) \log(1-p-q) - p \log p }{1+p+2q} \\
  & \qquad - \frac{q \log q  + (1-p) \log( 1-p )  }{1+p+2q} \\
  & \qquad + \log(1+p+2q)
    ~.
\end{align*}
This is the total future information that can possibly be predicted using past
observations. The structure of \emph{how} this information is unraveled over
time is revealed in the excess entropy spectrum:
\begin{align*}
\mathcal{E}(\omega) & =   \pi \hmu \delta(\omega) \\ 
  & \qquad +  \sum_{m=0}^{\nu_0 - 1} \cos \bigl( (m+1) \, \omega \bigr)
  \StartMS W_0 W^m \HWA \\
  & \qquad + \sum_{\lambda \in \Lambda_W \setminus 0}
  \sum_{m=0}^{\nu_\lambda - 1} 
  \text{Re} \biggl(
  \frac{ \StartMS W_{\lambda, m} \HWA }{(e^{i \omega} -\lambda)^{m+1}}
  \biggr) \\ 
  & = \pi \hmu \delta(\omega) + \frac{2q}{1+p+2q} \cos ( 2 \omega ) \\
  & \qquad + \bigl( \EE
  + \frac{ p \log p - (1+p) \log (1+p) -2q }{ 1+p+2q } \bigr)
  \cos ( \omega ) \\
  & \qquad + \frac{-p \log p  + (1+p) \log (1+p) }{1+p+2q}
  \text{Re} \biggl( \frac{ e^{i \omega} - p }{e^{i 2 \omega} - p} \biggr) \\
  & \qquad + \tfrac{2p }{1+p+2q}
  \text{Re} \biggl( \frac{ 1 - e^{i \omega} }{e^{i 2 \omega} - p}   \biggr)
  ~.
\end{align*}
From this, we observe that $\EE = \lim_{\omega \to 0} \mathcal{E}(\omega)$.

\subsection{Synchronizing to predict optimally}

To analyze the information-processing cost of synchronizing to a process, we
need its \eM\ \syncMSP. We already constructed this in the last section.
Hence, we can immediately evaluate the resources necessary for synchronizing to
and predicting the $(2\text{-}1)$-GP-$(2)$ Process.

The only novel piece needed for this is the vector of mixed-state entropies $\Hmxst$. A simple calculation yields:
\begin{align*}
\Hmxst = \begin{bmatrix}
\log(1+p+2q) - \tfrac{p \log p + 2q \log q}{1+p+2q} \\
1 \\
\log(1+p) - \tfrac{p \log p}{1+p}  \\
1\\
0 \\
0 \\
0 \\
0 
\end{bmatrix}
  ~,
\end{align*}
where $\log$ is again understood to be the base-2 logarithm $\log_2$.

An observer, tasked with predicting the future as well as possible, must
synchronize to the causal state of the dynamic. During the metadynamics of
synchronization, the observer on average will pick up synchronization
information according to the remaining causal-state uncertainty
$\mathcal{H}^+(L)$ after an observation interval of $L$ steps:
\begin{align*}
& \mathcal{H}^+(L) \\
& = \sum_{\lambda \in \Lambda_\mathcal{W} \atop \lambda \neq 0}
  \sum_{m = 0}^{\nu_\lambda -1} 
  \StartMS \mathcal{W}_{\lambda, m} \Hmxst
  \binom{L}{m} \lambda^{L-m} \\ 
  &  \qquad +  \left[ 0 \in \Lambda_\mathcal{W} \right] 
	\sum_{m=0}^{\nu_0 - 1} \delta_{L, m} 
	\StartMS \mathcal{W}_0 \mathcal{W}^m \Hmxst \\
  & = \delta_{L, 0} \bigl[ \log(1+p+2q)
  - \frac{2q \log q + (1+p) \log (1+p)}{1+p+2q} \bigr] \\
  & \qquad +  \delta_{L, 1} \tfrac{2q}{1+p+2q} 
  + \frac{ \langle \sqrt{p} \Hmxst + (-1)^L \langle ^- \! \sqrt{p} \Hmxst }
  {1+p+2q} \, p^{L/2}
 ~.
\end{align*}
More explicitly, for $L \geq 2$, this becomes:
\begin{align*}
\mathcal{H}^+(L) & = 
	\begin{cases}
	\frac{ 2 \sqrt{p} }{1+p+2q} \, p^{L/2} & \text{ \small for odd } L \\
	\frac{ (1+p) \log(1+p) -p \log p }{ 1+p+2q} \, p^{L/2} & \text{ \small for even } L
	\end{cases}
  ~.
\end{align*}

The total synchronization information accumulated is then:
\begin{align*}
\SI & = 
\!\!\! \sum_{\lambda \in \Lambda_\mathcal{W} \setminus \{ 1 \} } 
    \sum_{m=0}^{\nu_\lambda - 1} 
    \frac{1}{(1-\lambda)^{m+1}} 
    \StartMS  \mathcal{W}_{\lambda, m}  \Hmxst \\
  & =  \frac{ 2q (1 - \log q) }{ 1+p+2q }  
    + \frac{ p \bigl[ 2 - \log p + (1+p) \log (1+p) \bigr] }
	{ (1-p)(1+p+2q) } \\    
  & \qquad + \log(1+p+2q) 
  ~.
\end{align*}

Even after synchronization, an observer must update an average of $\bmu$ of its bits of information per observation and must keep track of a net $\Cmu$ bits of information to stay synchronized, where:
\begin{align*}
\Cmu & = \H[\pi] \\
     & = \langle \delta_\pi  \Hmxst \\ 
  & = \log(1+p+2q) - \frac{p \log p + 2q \log q}{1+p+2q} ~.
\end{align*}

An interesting feature of prediction is a process' crypticity:
\begin{align*}
\PC & = \Cmu - \EE \\
    & = \frac{3q \log q + (1-p) \log (1-p)}{1+p+2q} \\\
    & \qquad - \frac{(1-p-q) \log (1-p-q) }{1+p+2q}
	~,
\end{align*}
This gives, as a function of $p$ and $q$, the minimal overhead of additional
memory about the past---beyond the information that the future can possibly
share with it---that must be stored for optimal prediction.

In summary, we now more fully appreciate, via this rather complete analysis, the
fundamental limits on predictability of our example stochastic process.  It
showed many of the qualitative features, both in terms of the calculation and
system behavior, that should be expected when analyzing prediction, based on
the more general results of the main development. The procedures can be
commandeered to apply to any inference algorithm that yields a generative
model, whether classical machine learning, Bayesian Structural
Inference~\cite{Stre13a}, or any other favorite inference tool. With a
generative model in hand, synchronizing to real world data---necessary to make
good predictions about the real world's future---follows the \syncMSP\
metadynamics. The consequences for prediction is that typically there will be a
finite epoch of symmetry collapse followed by a slower asymptotic
synchronization that allows improved prediction, as longer observations induce
a refined knowledge of what lies hidden.

\bibliography{chaos}

\end{document}